**Title: Critical weaknesses in shielding strategies for COVID-19**

**Authors:** Cameron A. Smith[1], Christian A. Yates[1]†, Ben Ashby[1,2]†*

**Affiliations:**

[1]Department of Mathematical Sciences, University of Bath; Bath, BA2 7AY, UK.

[2]Milner Centre for Evolution, University of Bath; Bath, BA2 7AY, UK.

*Corresponding author. Email: bna24@bath.ac.uk.

†Equal contribution

**Abstract:**

The COVID-19 pandemic, caused by the coronavirus SARS-CoV-2, has led to a wide range of non-pharmaceutical interventions being implemented around the world to curb transmission. However, the economic and social costs of some of these measures, especially lockdowns, has been high. An alternative and widely discussed public health strategy for the COVID-19 pandemic would have been to 'shield' those most vulnerable to COVID-19 (minimising their contacts with others), while allowing infection to spread among lower risk individuals with the aim of reaching herd immunity. Here we retrospectively explore the effectiveness of this strategy using a stochastic SEIR framework, showing that even under the unrealistic assumption of perfect shielding, hospitals would have been rapidly overwhelmed with many avoidable deaths among lower risk individuals. Crucially, even a small (20%) reduction in the effectiveness of shielding would have likely led to a large increase (>150%) in the number of deaths compared to perfect shielding. Our findings demonstrate that shielding the vulnerable while allowing infections to spread among the wider population would not have been a viable public health strategy for COVID-19 and is unlikely to be effective for future pandemics.

## §1: Introduction

The COVID-19 pandemic has caused unprecedented health, economic, and societal challenges. As of February 2022, around 400 million cases and more than 5.5 million deaths have been confirmed, although the true numbers are thought to be far higher (1). Prior to (and during) the rollout of vaccines, most countries introduced a range of non-pharmaceutical interventions (NPIs) to bring infections under control, including social distancing, travel restrictions, and lockdowns. While the effectiveness of different NPIs has varied within and between populations and over time, they have been largely effective at bringing outbreaks under control (2–4). A widely discussed alternative approach would have been to limit most NPIs to the most vulnerable subpopulations while allowing those at lower risk to live with few or no restrictions (4–6). 'Shielding' (or 'focused protection'), appeared to offer the possibility of avoiding the various societal costs of universal NPIs by leveraging the uneven risk profile of COVID-19, which is heavily skewed towards the elderly and those with certain pre-existing conditions (7,8). In theory, by allowing infections to spread with little to no suppression among the lower-risk population during a temporary shielding phase, the higher-risk population would subsequently be protected by herd immunity (9). Several countries either openly or reportedly embraced this strategy during the early stages of the pandemic. Sweden, for example, chose to impose few restrictions on the general population while banning visits to long-term care (LTC) facilities (10), and the UK initially appeared to opt for a shielding strategy (11) before implementing a national lockdown. In the autumn of 2020, many countries experienced a resurgence in infections following the lifting of NPIs, leading to a renewed debate about the merits of shielding, driven by the Great Barrington Declaration which called for "focused protection of older people and other high-risk groups" while allowing uncontrolled viral transmission among lower-risk individuals (12,13).

It is important to retrospectively assess the feasibility of shielding as a public health strategy, not only for public inquiries into COVID-19 and future pandemic preparedness, but also for countries where levels of vaccination remain low. Moreover, new variants may emerge which substantially escape vaccine-induced immunity, thus requiring a renewed choice between lockdowns and shielding while vaccines are updated. Although superficially appealing, serious practical and ethical concerns have been raised about shielding as a strategy to mitigate the impact of COVID-19 (14). Yet there has been little mathematical modelling to determine the effectiveness of shielding under realistic conditions (4–6). Crucially, the combined consequences of imperfect shielding, uneven distributions of immunity, and changes in contact behaviour among lower-risk individuals have yet to be explored.

Here, we use a mathematical model to evaluate whether shielding the most vulnerable while allowing infections to spread among lower-risk members of the population would have been an effective strategy to combat COVID-19. Our simulations are intended as illustrative examples of how shielding would have likely performed during the early stages of the pandemic, with the aim of informing future pandemic preparedness. We employ a stochastic SEIR model (see §2.1 and Fig. 1) where the population is structured by risk of mortality (higher or lower risk) and location (community or LTC facilities). Our model is loosely based on an idealized large city in England (although our qualitative results would apply to similar countries) consisting of 1 million people, 7% of whom are at higher-risk of mortality from COVID-19, with 10% of higher-risk individuals situated in LTC facilities (15,16). We compare epidemics under no shielding, with imperfect (partial reduction in contacts for higher-risk individuals) and perfect shielding (no contacts for higher-risk individuals), with shielding restrictions lifted when cases fall below a given threshold (see §2.1).

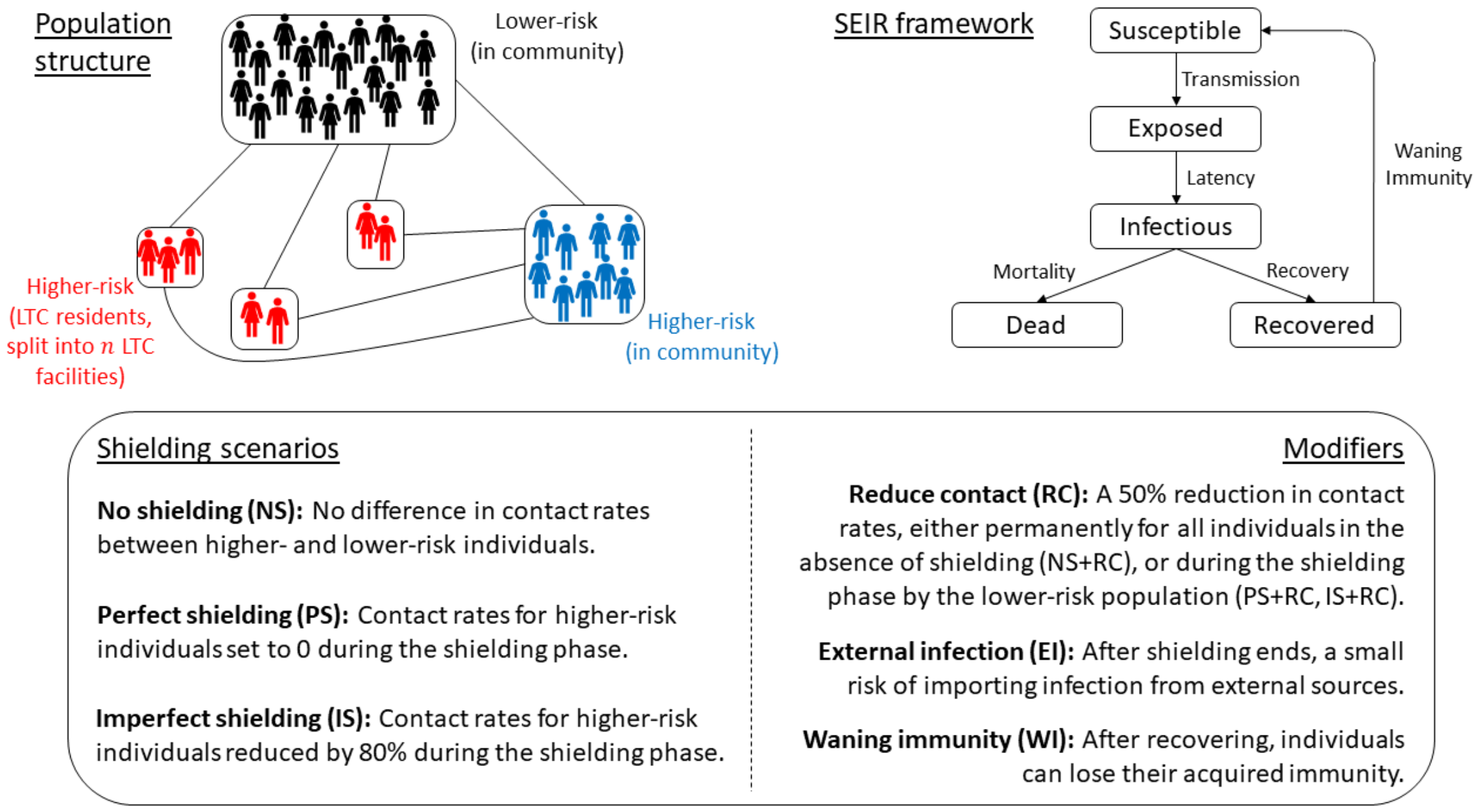

**Fig. 1. Schematic for the model.** Top left: Population structure, with lines indicating contacts between subpopulations, each of which are well-mixed. Top right: Transitions through the different infection states in the SEIR model. Bottom: Description of the shielding scenarios and modifying assumptions.

This paper is arranged as follows. In §2 we introduce the modelling framework and the methods that we use: §2.1 contains the model formulation, §2.2 the calculation of the basic reproductive number for our model, and §2.3 the calculation of hospitalisation rates. In §3 we present our results, and we discuss our findings in §4.

## §2: Materials and Methods

### §2.1: Model formulation

We simulate the spread of COVID-19 through the population of a large hypothetical city in England ($N = 1{,}000{,}000$). We consider a closed population (no births, non-disease related death or immigration) that is divided into three groups: a proportion $h$ of higher-risk individuals, with a proportion $ch$ of those living in the community $(H_C)$ and $(1-c)h$ living in $n$ long-term care (LTC) facilities $\left(H_F^i\right)$ (for $i = 1, \ldots, n$), with the remaining fraction of the total population, $1-h$, being lower-risk individuals living in the community $(L)$. We define the number of people in each of the subpopulations to be $N_L, N_{H_C}$ and $N_{H_F^i}$ for the lower-risk community, higher-risk community, and long-term care residents in facility $i$ (for $i = 1, \ldots, n$), respectively. Using approximate figures for those classed as clinically extremely vulnerable (CEV) in England, we set $h = 0.07$ (7% at higher risk of mortality from COVID-19) and $c = 0.897$ (around 90% of higher-risk individuals live in the community). To reflect variation in the sizes of LTC facilities in England, we assume that LTC residents are distributed evenly over small, medium and large facilities: there are 120 small LTC facilities, each with 20 residents; 48 medium LTC facilities, each with 50 residents each, and 24 large LTC facilities, each of which house 100 residents. This gives an average of 37.5 residents per LTC facility, which is close to the UK average 39 (17).

We define the infection fatality ratio (IFR) for those at lower risk to be $\alpha_L$, and for those at higher risk to be $\alpha_H$, with $(\alpha_L < \alpha_H)$. While many studies consider age stratified IFRs (7,8,18), there is comparatively little data on the IFR for CEV individuals. We assume that the lower-risk group consists of healthy people who are generally younger. IFR estimates for younger age groups range from 0.000097 (8) for the under 25s to 0.0052 (8) or 0.0094 (7) for the 45-64 age group. Conservatively, we choose a value towards the lower end of these estimates at 0.001 for the lower-risk group. For the higher-risk populations, we need to consider not only age, but also risk factors associated with being CEV. The majority of LTC residents are elderly, and so we look at the IFR for elderly populations as a proxy for this group. IFRs for 75-year-olds and over have been estimated to be as high as 0.1164 (7) and 0.147 (8), and previous modelling has assumed IFRs of 0.051 and 0.093 for the 70-79 and 80+ age groups, respectively (19). We again choose a conservative estimate, setting the IFR for higher-risk individuals both in the community and in LTC facilities at 0.05. Our choice of IFRs are only approximations with the intention of illustrating how different shielding scenarios affect changes in cumulative deaths. Other reasonable choices of IFR for the different risk subcategories in our model do not qualitatively affect our conclusions. We set the average incubation period $(1/\sigma)$ to be 5 days (19) and the average infectious period $(1/\Gamma)$ to be 2 days (19), which are assumed to be the same for all individuals, and the basic reproduction number, $R_0$, to be 3 (20) (see §2 for derivation). These parameters yield an unmitigated doubling time of around 3.2 days. For further simulations with $R_0 = 2.5$ and $R_0 = 3.5$, together with a sensitivity analysis for other key parameters, please see §S1 of the *Supplementary material*. When included, we use a 240-day average period for waning immunity $(1/\nu)$, which is suggested to be an upper limit to the natural immunity conferred from SARS-CoV-2 according to the World Health Organisation (21).

Each individual in the population is assigned one of five epidemiological states: $S$ for those susceptible to the disease, $E$ for those exposed but not yet infectious, $I$ for those infected and able to transmit the disease, $R$ for those who have recovered (recovery is assumed to lead to full lifelong immunity), and $D$ for those who have died from the disease. We then define $S_i, E_i, I_i, R_i$ and $D_i$ for $i \in \{L, H_C, H_F^1, \dots, H_F^n\}$ to be the total number of susceptible, exposed, infected, recovered, and dead individuals in subpopulation $i$. Susceptible individuals can become exposed through two pathways. Firstly, they may be "externally" infected from a member outside the population (for example, from another city or country), which we assume occurs at a rate $\eta_i(t)$ for subpopulation $i$ at time $t$. Alternatively, infected individuals of type $j$ may transmit the disease to susceptible individuals of type $i$ with rate $\beta_{ij} = \beta_0 r p_{ij}$, where $\beta_0$ is the transmission probability per contact, $r$ is the average number of contacts in the absence of restrictions, and $p_{ij}$ is the proportion of contacts that a person of type $i$ has with a person of type $j$. Written as a transmission matrix $\boldsymbol{\beta}$, we have:

$$\boldsymbol{\beta} = \beta_0 r \begin{pmatrix} p_{LL} & p_{LH_C} & p_{LH_F^1} & \cdots & p_{LH_F^n} \\ p_{H_C L} & p_{H_C H_C} & p_{H_C H_F^1} & \cdots & p_{H_C H_F^n} \\ p_{H_F^1 L} & p_{H_F^1 H_C} & p_{H_F^1 H_F^1} & \cdots & p_{H_F^1 H_F^n} \\ \vdots & \vdots & \vdots & \ddots & \vdots \\ p_{H_F^n L} & p_{H_F^n H_C} & p_{H_F^n H_F^1} & \cdots & p_{H_F^n H_F^n} \end{pmatrix}. \tag{1}$$

We assume that there is no direct contact between each of the LTC facilities, that a proportion $\lambda$ of an LTC resident's contacts occur within the same LTC facility, and that this proportion is the same for all LTC facilities. This yields:

$$p_{H_F^i H_F^j} = \lambda \delta_{i,j}, \tag{2}$$

where $\delta_{i,j}$ is the Kronecker delta, which takes the value 1 if $i = j$ and is 0 otherwise. The remainder of an LTC resident's contacts will occur with individuals in the community, normalised by the proportion of the population that is in the community:

$$p_{H_F^i L} = \frac{(1-h)(1-\lambda)}{1-h(1-c)}, \tag{3a}$$

$$p_{H_F^i H_C} = \frac{ch(1-\lambda)}{1-h(1-c)}, \tag{3b}$$

which holds for every $i \in \{1, \ldots, n\}$. The proportion of contacts that the lower- and higher-risk communities have with each care home is calculated as follows:

$$p_{LH_F^i} = \frac{N_{i+2}}{N} \frac{p_{H_F^i L}}{1-h}, \tag{4a}$$

$$p_{H_C H_F^i} = \frac{N_{i+2}}{N} \frac{p_{H_F^i H_C}}{hc}, \tag{4b}$$

where we have multiplied the contact rate in the opposite direction by the proportion of individuals that live in LTC facility $i$, and divide through by the proportion in each community group. We can calculate the remaining values in a similar way to $p_{H_F^i L}$ and $p_{H_F^i H_C}$ (i.e. by multiplying the remaining $1 - p_{H_F^i L}$ (and $1 - p_{H_F^i H_C}$) by the corresponding proportion of each subpopulation in the community), yielding the following transmission matrix:

$$\boldsymbol{\beta} = \beta_0 r \begin{pmatrix} \frac{(1-h)\left(1-\sum_i p_{LH_F^i}\right)}{1-h(1-c)} & \frac{ch\left(1-\sum_i p_{LH_F^i}\right)}{1-h(1-c)} & p_{LH_F^1} & \cdots & p_{LH_F^n} \\ \frac{(1-h)\left(1-\sum_i p_{H_C H_F^i}\right)}{1-h(1-c)} & \frac{ch\left(1-\sum_i p_{H_C H_F^i}\right)}{1-h(1-c)} & p_{H_C H_F^1} & \cdots & p_{H_C H_F^n} \\ p_{H_F^1 L} & p_{H_F^1 H_C} & \lambda & \cdots & 0 \\ \vdots & \vdots & \vdots & \ddots & \vdots \\ p_{H_F^n L} & p_{H_F^n H_C} & 0 & \cdots & \lambda \end{pmatrix}. \tag{5}$$

We implement non-pharmaceutical interventions (NPIs) by multiplying the transition matrix $\boldsymbol{\beta}$ element-wise by a shielding matrix $\boldsymbol{Q}$, whose entries lie between 0 and 1. A value of $Q_{ij} = 1$ denotes that interventions, if any, do not impact the contact rates between subpopulation $i$ and subpopulation $j$, so that contacts between the two occur as normal, while a value of $Q_{ij} = 0$ ceases all contacts between subpopulations $i$ and $j$. We further assume that the interventions are symmetric, so that $Q_{ij} = Q_{ji}$. The matrix $\boldsymbol{Q}$ is characterised by six different values and takes the following form:

$$\boldsymbol{Q} = \begin{pmatrix} q_1 & q_4 & q_5 & \cdots & q_5 \\ q_4 & q_2 & q_6 & \cdots & q_6 \\ q_5 & q_6 & q_3 & \cdots & 0 \\ \vdots & \vdots & \vdots & \ddots & \vdots \\ q_5 & q_6 & 0 & \cdots & q_3 \end{pmatrix}. \tag{6}$$

We consider three different shielding scenarios: no shielding (NS), imperfect shielding (IS), and perfect shielding (PS). In the imperfect and perfect shielding scenarios, shielding begins at the start of each simulation and ends once incidence falls below a threshold of 60 new cases per 100,000 in the population per week. This threshold is chosen based on the number of new cases recorded in the UK on the 1st April 2021 when shielding advice ended (22). Once shielding ends it does not start again if cases rise. Coupled to each of these shielding scenarios, we include modifiers: reduced contact of the population during the shielding phase (RC) and the addition of an external force of infection into the community post shielding (EI). The RC modifier is motivated by Apple mobility data, which shows that in the week leading up to the first full lockdown in England on the 23rd March 2020, levels of movement may have dropped by around 70% (23), as measured by the number of requests for directions using Apple Maps. This indicates that members of the population may voluntarily reduce their contact when faced with an emerging pandemic. When there is no shielding strategy and the RC modifier is applied, we assume that all subpopulations reduce their contact equally. When either perfect or imperfect shielding is applied, the reduced contact is applied to the lower-risk population only as the higher-risk population already has reduced contact due to shielding. The second modifier, which introduces external infections, is motivated by those who enter the population and have the potential to infect those in the focal population. We assume that during the shielding phase, the population is closed, and so this modifier is only applied after shielding ends. Table S1 shows values for the entries of $\boldsymbol{Q}$ for the nine scenarios in §3.

To evolve the model system, we employ the Gillespie stochastic simulation algorithm (SSA) (24) (The code for our implementation can be found online (25)). The transition rates between the different states of the system are defined using the numbers of individuals in each of the subpopulations. Let $\underline{Y}_i^t = (S_i, E_i, I_i, R_i, D_i)$ be the state variable for the $i^{\text{th}}$ subpopulation at time $t$. Then we can define the transition probabilities between states over a small time interval $(t, t + \delta t)$ to be as follows:

$$\mathbb{P}\left(\underline{Y_i}^{t+\delta t} - \underline{Y_i}^{t} = (-1,1,0,0,0)\right) = \left(\eta_i(t)S_i + \sum_j \frac{Q_{ij}\beta_{ij}S_i I_j}{N_j}\right)\delta t, \tag{7a}$$

$$\mathbb{P}\left(\underline{Y_i}^{t+\delta t} - \underline{Y_i}^{t} = (0,-1,1,0,0)\right) = \sigma E_i \delta t, \tag{7b}$$

$$\mathbb{P}\left(\underline{Y_i}^{t+\delta t} - \underline{Y_i}^{t} = (0,0,-1,1,0)\right) = \Gamma I_i(1-\alpha_i)\delta t, \tag{7c}$$

$$\mathbb{P}\left(\underline{Y_i}^{t+\delta t} - \underline{Y_i}^{t} = (0,0,-1,0,1)\right) = \Gamma I_i \alpha_i \delta t, \tag{7d}$$

$$\mathbb{P}\left(\underline{Y_i}^{t+\delta t} - \underline{Y_i}^{t} = (1,0,0,-1,0)\right) = \nu \delta t, \tag{7e}$$

which holds for every subpopulation $i$, whilst holding each of the other $\underline{Y_j}^{t}$s constant for every $j \neq i$. Each of the above probabilities is associated with the transition of an individual between successive disease states. The first is the conversion of a susceptible to an exposed individual through coming into contact with an infected individual from any of the other subpopulations $j \in \{L, H_C, H_F^1, \dots, H_F^n\}$. Note that the first term in the bracket is the external infection term, which is always 0 during the shielding phase, and when included, is non-zero only in the lower-risk and higher-risk subpopulations in the community. The second characterises the transition from being exposed to infectious, the third the recovery of an infected individual and the fourth the death of an infected individual. We run each of our stochastic simulations for 600 days and calculate averages and variances over 100 independent repeats, initialised with 10 lower-risk individuals in the infected class. We remove any instances of immediate stochastic die out from our analysis (this is a rare occurrence).

### §2.2: The basic reproduction number

To approximate the basic reproduction number for our simulation, we employ the mean-field ordinary differential equations (ODEs) for each subpopulation, for which we combine all LTC facility residents into one large subpopulation. In this section, we assume that the elements $\beta_{ij}$ contain the appropriate shielding matrix term. The mean-field equations (assuming no interventions) are:

$$\frac{dS_L}{dt} = -\frac{\beta_{LL}S_L I_L}{N_L} - \frac{\beta_{LH_C}S_L I_{H_C}}{N_{H_C}} - \frac{\beta_{LH_F}S_L I_{H_F}}{N_{H_F}}, \tag{8a}$$

$$\frac{dS_{H_C}}{dt} = -\frac{\beta_{H_CL}S_{H_C} I_L}{N_L} - \frac{\beta_{H_CH_C}S_{H_C} I_{H_C}}{N_{H_C}} - \frac{\beta_{H_CH_F}S_{H_C} I_{H_F}}{N_{H_F}}, \tag{8b}$$

$$\frac{dS_{H_F}}{dt} = -\frac{\beta_{H_FL}S_{H_F} I_L}{N_L} - \frac{\beta_{H_FH_C}S_{H_F} I_{H_C}}{N_{H_C}} - \frac{\beta_{H_FH_F}S_{H_F} I_{H_F}}{N_{H_F}}, \tag{8c}$$

$$\frac{dE_L}{dt} = \frac{\beta_{LL}S_L I_L}{N_L} + \frac{\beta_{LH_C}S_L I_{H_C}}{N_{H_C}} + \frac{\beta_{LH_F}S_L I_{H_F}}{N_{H_F}} - \sigma E_L, \tag{8d}$$

$$\frac{dE_{H_C}}{dt} = \frac{\beta_{H_CL}S_{H_C} I_L}{N_L} + \frac{\beta_{H_CH_C}S_{H_C} I_{H_C}}{N_{H_C}} + \frac{\beta_{H_CH_F}S_{H_C} I_{H_F}}{N_{H_F}} - \sigma E_{H_C}, \tag{8e}$$

$$\frac{dE_{H_F}}{dt} = \frac{\beta_{H_FL}S_{H_F} I_L}{N_L} + \frac{\beta_{H_FH_C}S_{H_F} I_{H_C}}{N_{H_C}} + \frac{\beta_{H_FH_F}S_{H_F} I_{H_F}}{N_{H_F}} - \sigma E_{H_F}, \tag{8f}$$

$$\frac{dI_L}{dt} = \sigma E_L - \Gamma I_L, \tag{8g}$$

$$\frac{dI_{H_C}}{dt} = \sigma E_{H_C} - \Gamma I_{H_C}, \tag{8h}$$

$$\frac{dI_{H_F}}{dt} = \sigma E_{H_F} - \Gamma I_{H_F}. \tag{8i}$$

The recovered and death classes have been omitted here because they are not required for the calculation. We employ the next generation matrix method (26) in order to find the basic reproduction number. We linearise the infected state ODEs ($E_i$ and $I_i$) in system (8) about the disease-free equilibrium

$$\underline{S}_0 = \left(S_L, S_{H_C}, S_{H_F}, E_L, E_{H_C}, E_{H_F}, I_L, I_{H_C}, I_{H_F}\right) = \left(N_L, N_{H_C}, N_{H_F}, 0,0,0,0,0,0\right), \tag{9}$$

by writing $\underline{x} = \left(E_L, E_{H_C}, E_{H_F}, I_L, I_{H_C}, I_{H_F}\right)^T$ (where the superscript $T$ denotes the transpose) and obtaining an ODE $\dot{\underline{x}} = \boldsymbol{A}\underline{x}$, where:

$$\boldsymbol{A} = \begin{pmatrix} -\sigma & 0 & 0 & \beta_{LL} & \beta_{LH_C}\dfrac{N_L}{N_{H_C}} & \beta_{LH_F}\dfrac{N_L}{N_{H_F}} \\ 0 & -\sigma & 0 & \beta_{H_CL}\dfrac{N_{H_C}}{N_L} & \beta_{H_CH_C} & \beta_{H_CH_F}\dfrac{N_{H_C}}{N_{H_F}} \\ 0 & 0 & -\sigma & \beta_{H_FL}\dfrac{N_{H_F}}{N_L} & \beta_{H_FH_C}\dfrac{N_{H_F}}{N_{H_C}} & \beta_{H_FH_F} \\ \sigma & 0 & 0 & -\Gamma & 0 & 0 \\ 0 & \sigma & 0 & 0 & -\Gamma & 0 \\ 0 & 0 & \sigma & 0 & 0 & -\Gamma \end{pmatrix}. \tag{10}$$

We split matrix (10) into components $\boldsymbol{T}$ and $\boldsymbol{\Sigma}$ which contain the transmission terms (or the terms relating to the mechanism by which individuals enter this truncated system) and all other terms respectively, so that:

$$\boldsymbol{T} = \begin{pmatrix} 0 & 0 & 0 & \beta_{LL} & \beta_{LH_C}\frac{N_L}{N_{H_C}} & \beta_{LH_F}\frac{N_L}{N_{H_F}} \\ 0 & 0 & 0 & \beta_{H_CL}\frac{N_{H_C}}{N_L} & \beta_{H_CH_C} & \beta_{H_CH_F}\frac{N_{H_C}}{N_{H_F}} \\ 0 & 0 & 0 & \beta_{H_FL}\frac{N_{H_F}}{N_L} & \beta_{H_FH_C}\frac{N_{H_F}}{N_{H_C}} & \beta_{H_FH_F} \\ 0 & 0 & 0 & 0 & 0 & 0 \\ 0 & 0 & 0 & 0 & 0 & 0 \\ 0 & 0 & 0 & 0 & 0 & 0 \end{pmatrix}, \tag{11}$$

$$\boldsymbol{\Sigma} = \begin{pmatrix} -\sigma & 0 & 0 & 0 & 0 & 0 \\ 0 & -\sigma & 0 & 0 & 0 & 0 \\ 0 & 0 & -\sigma & 0 & 0 & 0 \\ \sigma & 0 & 0 & -\Gamma & 0 & 0 \\ 0 & \sigma & 0 & 0 & -\Gamma & 0 \\ 0 & 0 & \sigma & 0 & 0 & -\Gamma \end{pmatrix}. \tag{12}$$

The next generation matrix $\boldsymbol{K}$ is then given by $\boldsymbol{K} = -\boldsymbol{T}\boldsymbol{\Sigma}^{-\boldsymbol{1}}$:

$$\boldsymbol{K} = -\boldsymbol{T}\boldsymbol{\Sigma}^{-1}$$

$$= \begin{pmatrix} \frac{\beta_{LL}}{\Gamma} & \frac{\beta_{LH_C}}{\Gamma}\frac{N_L}{N_{H_C}} & \frac{\beta_{LH_F}}{\Gamma}\frac{N_L}{N_{H_F}} & \frac{\beta_{LL}}{\Gamma} & \frac{\beta_{LH_C}}{\Gamma}\frac{N_L}{N_{H_C}} & \frac{\beta_{LH_F}}{\Gamma}\frac{N_L}{N_{H_F}} \\ \frac{\beta_{H_CL}}{\Gamma}\frac{N_{H_C}}{N_L} & \frac{\beta_{H_CH_C}}{\Gamma} & \frac{\beta_{H_CH_F}}{\Gamma}\frac{N_{H_C}}{N_{H_F}} & \frac{\beta_{H_CL}}{\Gamma}\frac{N_{H_C}}{N_L} & \frac{\beta_{H_CH_C}}{\Gamma} & \frac{\beta_{H_CH_F}}{\Gamma}\frac{N_{H_C}}{N_{H_F}} \\ \frac{\beta_{H_FL}}{\Gamma}\frac{N_{H_F}}{N_L} & \frac{\beta_{H_FH_C}}{\Gamma}\frac{N_{H_F}}{N_{H_C}} & \frac{\beta_{H_FH_F}}{\Gamma} & \frac{\beta_{H_FL}}{\Gamma}\frac{N_{H_F}}{N_L} & \frac{\beta_{H_FH_C}}{\Gamma}\frac{N_{H_F}}{N_{H_C}} & \frac{\beta_{H_FH_F}}{\Gamma} \\ 0 & 0 & 0 & 0 & 0 & 0 \\ 0 & 0 & 0 & 0 & 0 & 0 \\ 0 & 0 & 0 & 0 & 0 & 0 \end{pmatrix}. \tag{13}$$

The basic reproduction number, $R_0$, is given by the leading eigenvalue of $\boldsymbol{K}$. To simplify the resulting characteristic polynomial, we note that the matrix $\boldsymbol{\beta}$ takes the following form, calculated in an analogous way to §2.1 above:

$$\boldsymbol{\beta} = \begin{pmatrix} \beta_{LL} & \beta_{LH_C} & \beta_{LH_F} \\ \beta_{H_CL} & \beta_{H_CH_C} & \beta_{H_CH_F} \\ \beta_{H_FL} & \beta_{H_FH_C} & \beta_{H_FH_F} \end{pmatrix} = \begin{pmatrix} b_1q_1 & b_2q_4 & b_3q_5 \\ b_1q_4 & b_2q_2 & b_3q_6 \\ b_4q_5 & b_5q_6 & b_6q_3 \end{pmatrix}, \tag{14}$$

where:

$$b_1 = \beta_0 r \frac{(1-h)(1-b_3)}{1-h(1-c)}, \tag{15a}$$

$$b_2 = \beta_0 r \frac{ch(1-b_3)}{1-h(1-c)}, \tag{15b}$$

$$b_3 = \beta_0 r \frac{h(1-c)(1-\lambda)}{1-h(1-c)}, \tag{15c}$$

$$b_4 = \beta_0 r \frac{(1-h)(1-\lambda)}{1-h(1-c)}, \tag{15d}$$

$$b_5 = \beta_0 r \frac{ch(1-\lambda)}{1-h(1-c)}, \tag{15e}$$

$$b_6 = \beta_0 r \lambda, \tag{15f}$$

and the $q_i$s are as in equation (6). Substituting equations (15) into (13) and simplifying, we find the characteristic polynomial, $P(s)$, to be:

$$\begin{aligned} P(s) \\ = s^3 \Bigg[ s^3 - \frac{(b_1 q_1 + b_2 q_2 + b_6 q_3)}{\Gamma} s^2 \\ + \frac{b_6 q_3 (b_1 q_1 + b_2 q_2) - b_3 (b_4 q_5^2 + b_5 q_6^2) + b_1 b_2 (q_1 q_2 - q_4^2)}{\Gamma^2} s \\ - \frac{b_1 b_2 b_6 q_3 (q_1 q_2 - q_3 q_4) + b_1 b_3 b_5 q_6 (q_4 q_5 - q_1 q_6) + b_2 b_3 b_4 q_5 (q_4 q_6 - q_2 q_5)}{\Gamma^3} \Bigg]. \end{aligned} \tag{16a}$$

Under the no shielding scenario, $q_i = 1$ for every $i \in \{1, \dots, 6\}$ and we obtain the simplified characteristic polynomial:

$$P(s) = s^4 \left[ s^2 - \frac{1}{\Gamma}(b_1 + b_2 + b_6)s + \frac{1}{\Gamma^2}\big(b_6(b_1 + b_2) - b_3(b_4 + b_5)\big) \right]. \tag{16b}$$

Employing the fact that $b_1 + b_2 + b_3 = b_4 + b_5 + b_6 = \beta_0 r$, we obtain the form:

$$P(s) = s^4 \left(s - \frac{\beta_0 r}{\Gamma}\right)\left(s - \frac{b_1 + b_2 + b_6 - \beta_0 r}{\Gamma}\right). \tag{17}$$

The largest of the two eigenvalues that result from setting the characteristic polynomial to zero is $\beta_0 r/\Gamma$, and hence:

$$R_0 = \frac{\beta_0 r}{\Gamma}. \tag{18}$$

**§2.3: Hospitalisation**

To calculate the impact of our interventions on the occupancy of intensive care units (ICUs) during the epidemic, we utilise data from (19) on the age distribution of patients requiring ICU treatment (Table S3). These data allow us to calculate the probability that an individual who is infected requires ICU treatment. This probability is then *a posteriori* applied to our infection curve to estimate the numbers of people who would be in ICU. To calculate the probability of requiring ICU treatment given that an individual is symptomatic in the lower-risk group (assumed to be 66% of our infected class), we take a weighted average over all ages up to and including 64 (in a similar way to the calculation of IFR values), while the higher-risk subpopulations use ages 65 and over as a proxy. This yields probabilities of requiring ICU treatment for the lower- and higher-risk (community and LTC facilities) subpopulations, $\pi_L$ and $\pi_H$, of:

$$\pi_L = \sum_{\ell \in \mathcal{A}_L} pop_\ell \times ICU_\ell\,,$$

$$\pi_H = \sum_{\ell \in \mathcal{A}_H} pop_\ell \times ICU_\ell,$$

where $\mathcal{A}_L$ is the set of age categories below the age of 64, and $\mathcal{A}_H$ is the set of age categories above the age of 65. Also, $pop_\ell$ is the proportion of the population in the age category $\ell$, and $ICU_\ell$ is the probability that a symptomatic individual in age category $\ell$ is admitted to ICU. Each individual admitted to ICU is assumed to stay for ten days (on average) (19). The ICU capacity for the UK is approximately 8 ICU beds per 100,000 people (19).

**§3: Results**

An unmitigated epidemic with no shielding (NS) would have represented the worst-case scenario (Fig. 2, col. 1), with an estimated peak incidence of $4149.0 \pm 274.1$ (mean $\pm$ standard deviation) cases per 100,000 and a total of $415.1 \pm 6.5$ deaths per 100,000, equivalent to $230{,}795 \pm 3{,}615$ total deaths in England. This is likely a conservative estimate, as hospitals would have been rapidly overwhelmed, with intensive care unit (ICU) capacity exceeded by a factor of approximately 18 at the peak of the epidemic (Fig. 2D). In contrast, perfect shielding (PS) would have been the best-case scenario (although unattainable) (Fig. 2, col. 3), with a peak incidence of $3470.5 \pm 456.1$ cases per 100,000 but only $87.6 \pm 3.4$ deaths per 100,000. Perfect shielding represents a substantial improvement on an unmitigated epidemic (79% reduction in deaths), but almost all deaths would have been among lower-risk members of the population. In England, this would have equated to nearly 50,000 deaths among lower-risk individuals. As in the no shielding scenario, this is likely a conservative estimate as hospital capacity would have been rapidly overwhelmed: assuming an average duration of treatment of 10 days, ICU bed capacity in England would have been exceeded by over a factor of 10 at the peak of the epidemic with perfect shielding (see §2.3).

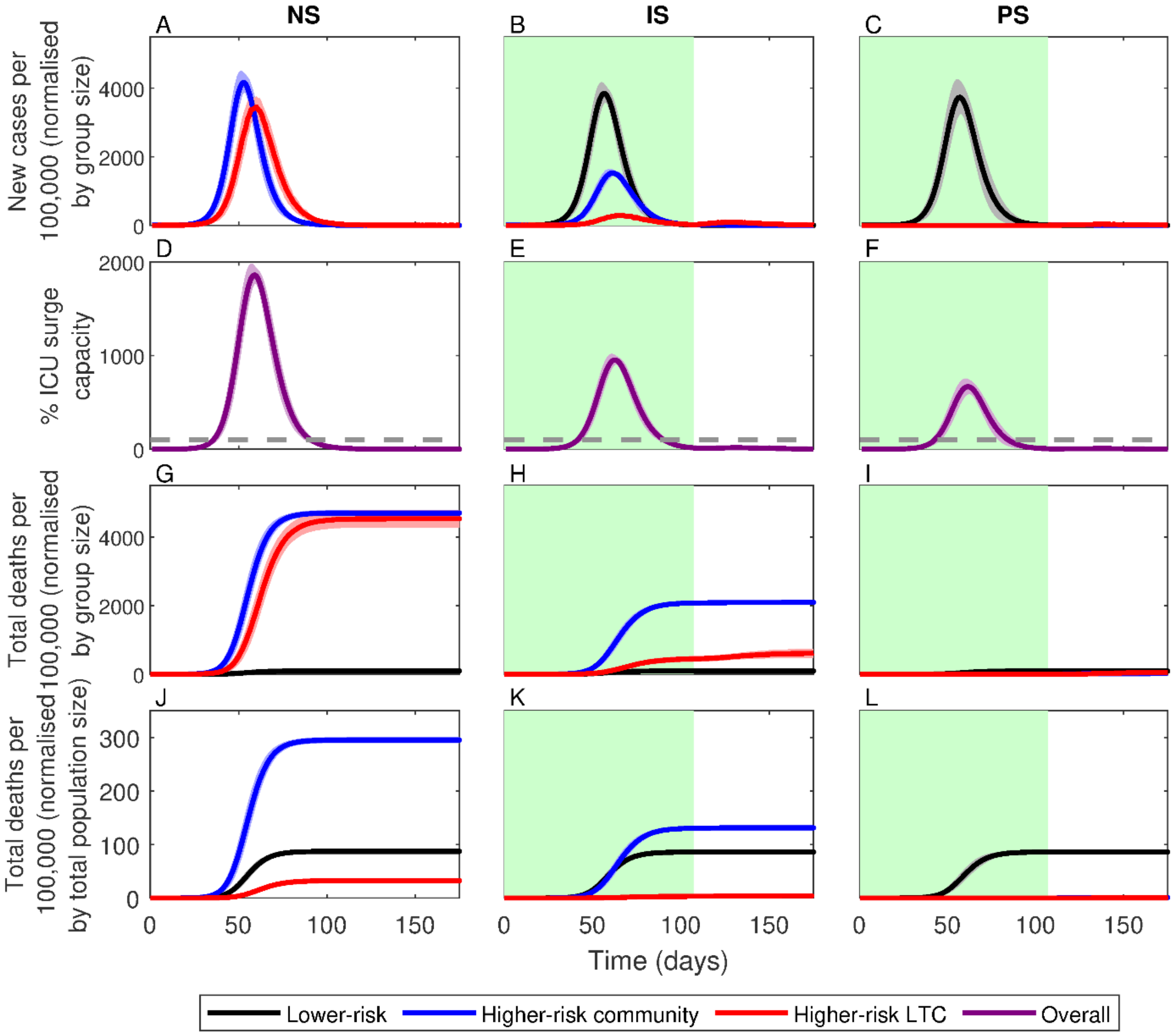

**Fig. 2. Simulations of no (NS), imperfect (IS) and perfect (PS) shielding.** Lines correspond to means for groups at: lower-risk (black), higher-risk in the community (blue) and in LTC facilities (red), with shading indicating ± 1 SD. Green shading indicates the shielding phase. Top row: daily number of new cases per 100,000 members of each group (i.e., new cases in each group multiplied by 100,000 and divided by group size). Second row: percentage of surge capacity ICU beds in demand (horizontal dashed line indicates full capacity). Third row: cumulative number of deaths per 100,000 members of each group (i.e., total deaths in each group multiplied by 100,000 and divided by group size). Bottom row: cumulative number of deaths per 100,000 members of the total population (i.e.,

total deaths multiplied by 100,000 and divided by total population size). Data averaged over 100 identically initialized stochastic repeats (see *Materials and methods*).

However, shielding would have been impossible to implement perfectly. LTC residents, for example, have contact with staff, and many higher-risk individuals in the community live with or receive care from lower-risk individuals. Between 14 May and 16 July 2020, only 58-63% of CEV people in England were able to follow guidelines to avoid contact completely (16), and despite strict restrictions on LTC facilities in Sweden and England during the first wave of the pandemic, a high proportion of COVID-19 deaths were LTC residents (15). Imperfect shielding, the first critical weakness of this strategy, therefore represents a more realistic scenario. If shielding had been only 80% effective while an otherwise unmitigated epidemic spread through the lower-risk population, we estimate that there would have been large outbreaks among higher-risk individuals both in the community and in LTC facilities (Fig. 2, col. 2) leading to a much higher death rate of $221.7 \pm 3.8$ per 100,000. Even a relatively small reduction in shielding effectiveness (20%) would have therefore led to a sharp increase in deaths (>150%) compared to perfect shielding (Fig. 3). Higher-risk individuals in the community would have been disproportionately affected due to imperfect shielding, with 200% higher death rates compared to LTC residents. Again, these figures are likely to be conservative as we estimate that hospital capacity would have been exceeded by a factor of 9.5 at the peak of the epidemic.

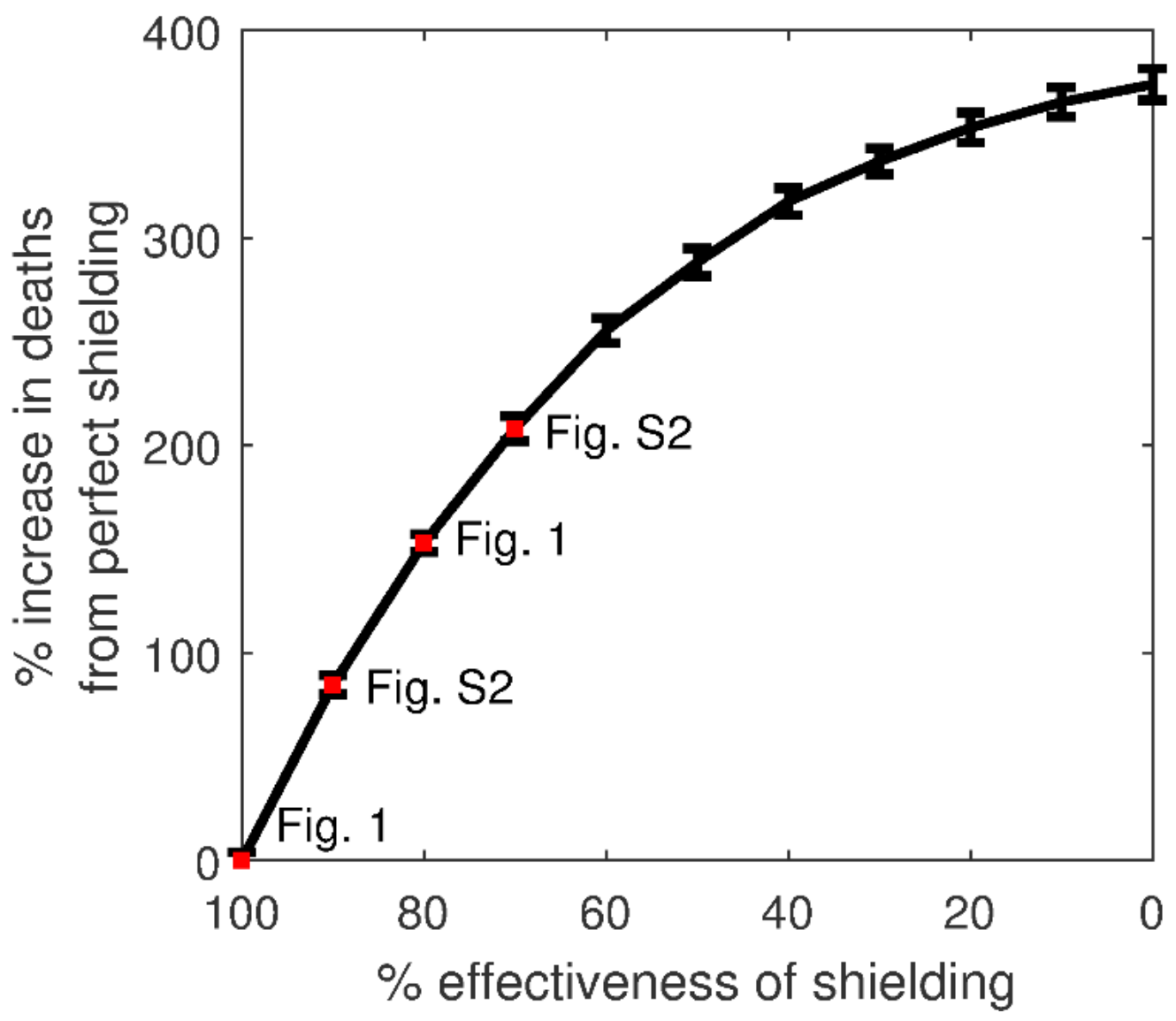

**Fig. 3. Relative deaths under varying levels of imperfect shielding (compared to perfect shielding).** Red dots and labels correspond to figures showing these scenarios, and vertical bars indicate ± 1 standard deviation.

The second critical weakness of the shielding strategy is that it relies on large numbers of lower-risk individuals becoming infected to build up immunity in the population. Yet many people would have likely changed their behaviour to avoid infection, leading to smaller, longer outbreaks with fewer infections and potentially leaving immunity levels below the threshold needed to prevent subsequent outbreaks (9). Prior to England's first national lockdown, mobility data shows that movement dropped by as much as 70% (23), and many people continued to take precautions, such as mask wearing and working from home, even after restrictions were fully lifted in July 2021 (28). A resurgence in cases leading to a second, deadlier wave, occurs in our modelling when reduced contact (50%) among lower risk individuals is combined with shielding, whether imperfect (IS+RC, 321.2 ± 11.5 deaths per 100,000) or perfect (PS+RC, 299.5 ± 7.5 deaths per 100,000) (Fig. 4, cols 2-3). Reduced contact among lower-risk individuals leads to much smaller peaks in incidence and hospitalizations, although ICU surge capacity would still likely have been exceeded without further restrictions (Fig. 4). It is also likely that people would have increasingly limited their contacts if healthcare services were overwhelmed, which would have further reduced the likelihood of reaching the herd immunity threshold before shielding ended.

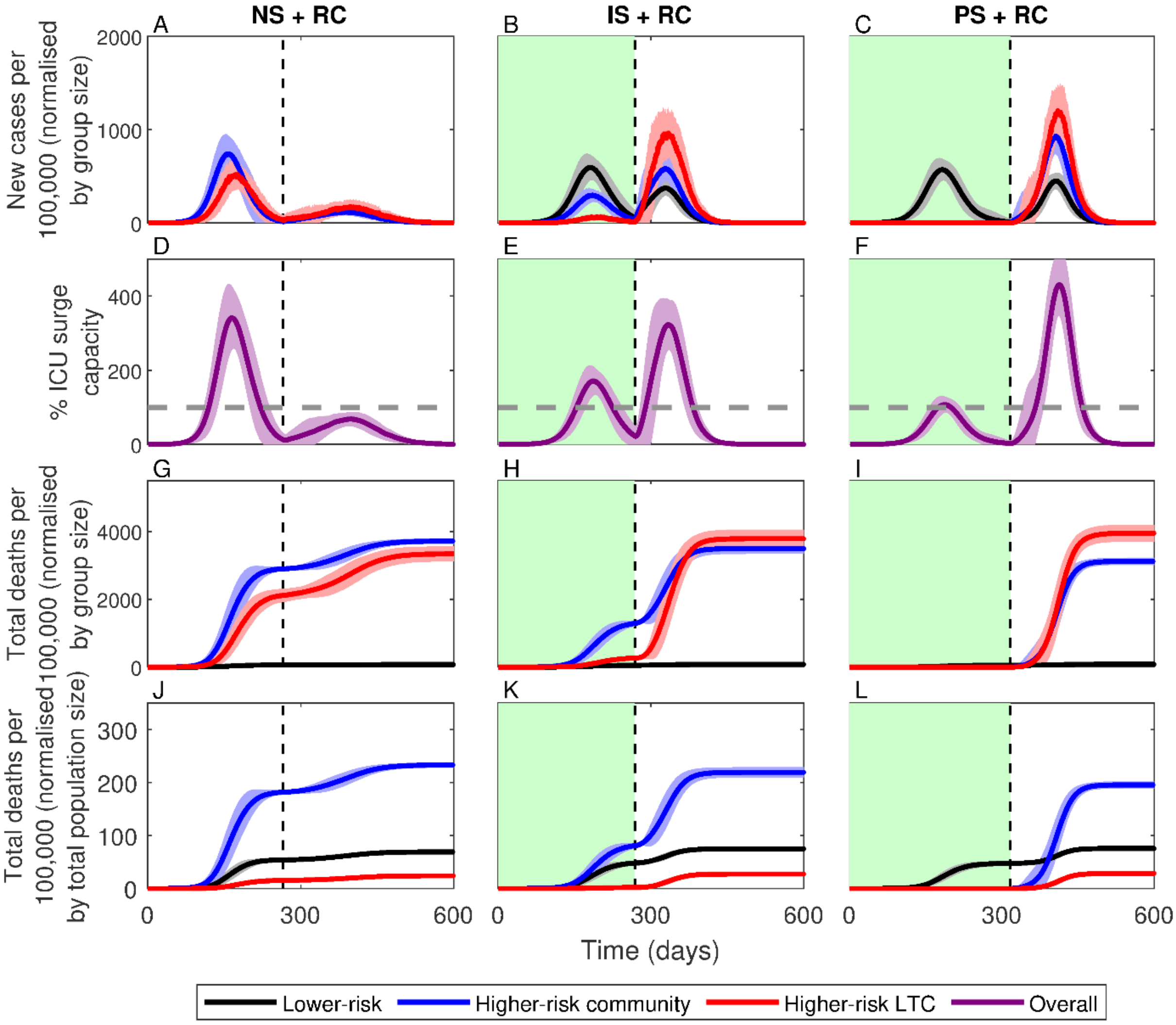

**Fig. 4. Simulations of the three shielding scenarios with 50% reduced contact (+RC).** RC occurs prior to the vertical dashed line. All other descriptions as in Fig. 2. Lines correspond to means for groups at: lower-risk (black), higher-risk in the community (blue) and in LTC facilities (red), with shading indicating ± 1 SD. Green shading indicates the shielding phase. NS: no shielding; IS: imperfect shielding; PS: perfect shielding.

A third critical weakness of the shielding strategy is that herd immunity only confers indirect protection and is only temporary. In theory, herd immunity would have been achieved primarily through infection of lower risk members of the population, conferring indirect protection to higher-risk individuals by preventing large outbreaks following the lifting of restrictions (See *Supplementary Material §S3* and Fig. S16). Yet many vulnerable members of the population would have remained at risk of infection after shielding had ended, from residual transmission in the community, from externally imported (EI) infections (e.g., due to international travel; Fig. 5) or from a resurgence in community transmission due to waning immunity (Fig. 6). Crucially, a heterogeneous distribution of immunity would have arisen in the population during the shielding phase, with LTC facilities remaining highly susceptible to local outbreaks once restrictions were lifted. If the shielding phase were to end prematurely while community transmission was still high or if infections were imported from other areas, local outbreaks would have likely still occurred in LTC facilities even if the population as a whole was above the herd immunity threshold. Similar effects have been observed for other pathogens, notably measles outbreaks in communities with low vaccination rates (29). In our simulations, we see that an external force of infection leads to a steady increase in deaths after shielding is lifted (Fig. 5), both among higher-risk individuals in the community despite herd-immunity being achieved, and among clusters of higher-risk individuals in LTC facilities in which herd immunity has not been achieved locally. Similarly, waning immunity leads to a resurgence in cases following the relaxation of shielding, leading to a substantial increase in deaths among those at greatest risk (Fig. 6).

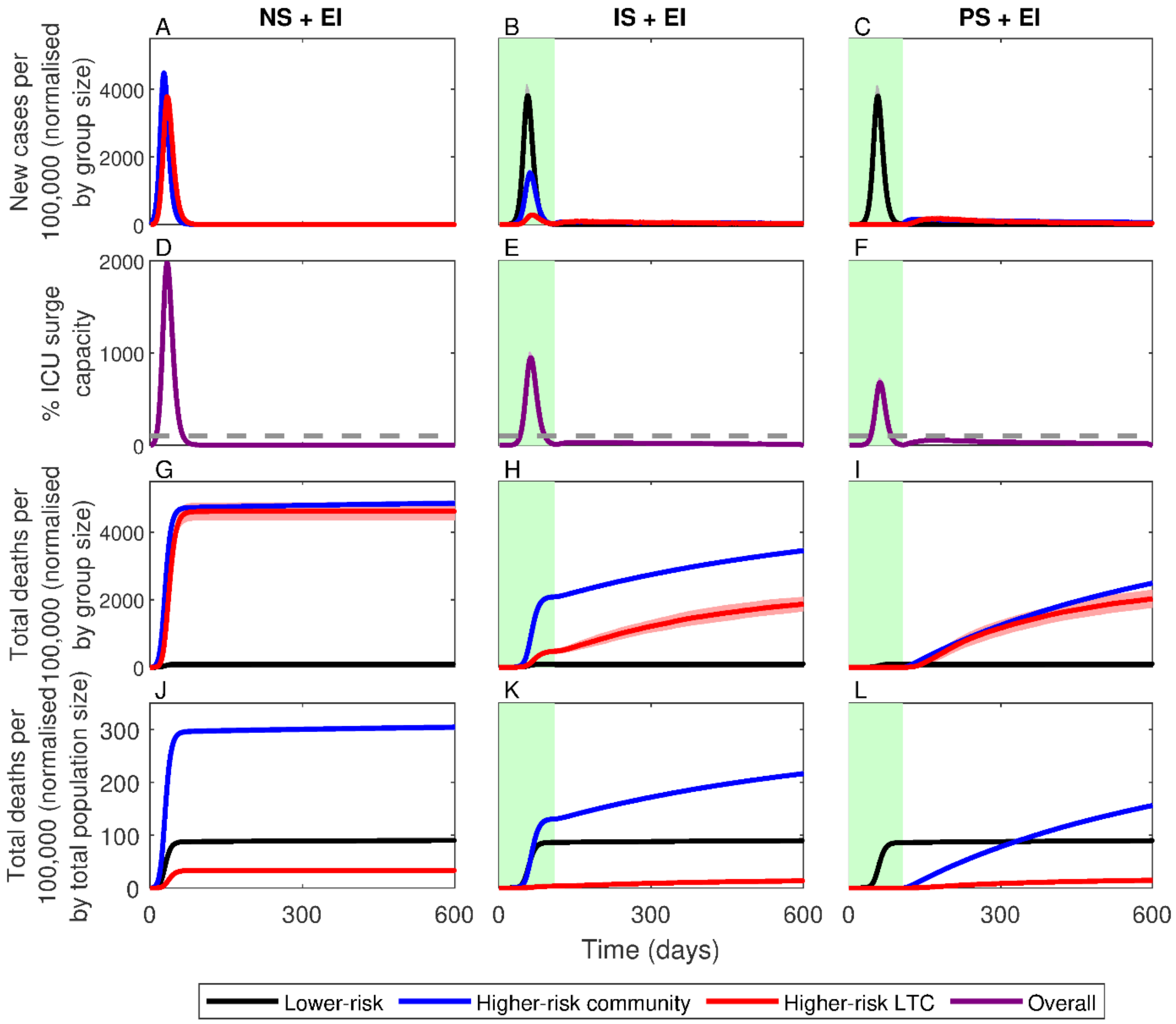

**Fig. 5. Simulations of the three shielding scenarios with external infections (+EI).** All other descriptions as in Fig. 2. Lines correspond to means for groups at: lower-risk (black), higher-risk in the community (blue) and in LTC facilities (red), with shading indicating ± 1 SD. Green shading indicates the shielding phase. NS: no shielding; IS: imperfect shielding; PS: perfect shielding.

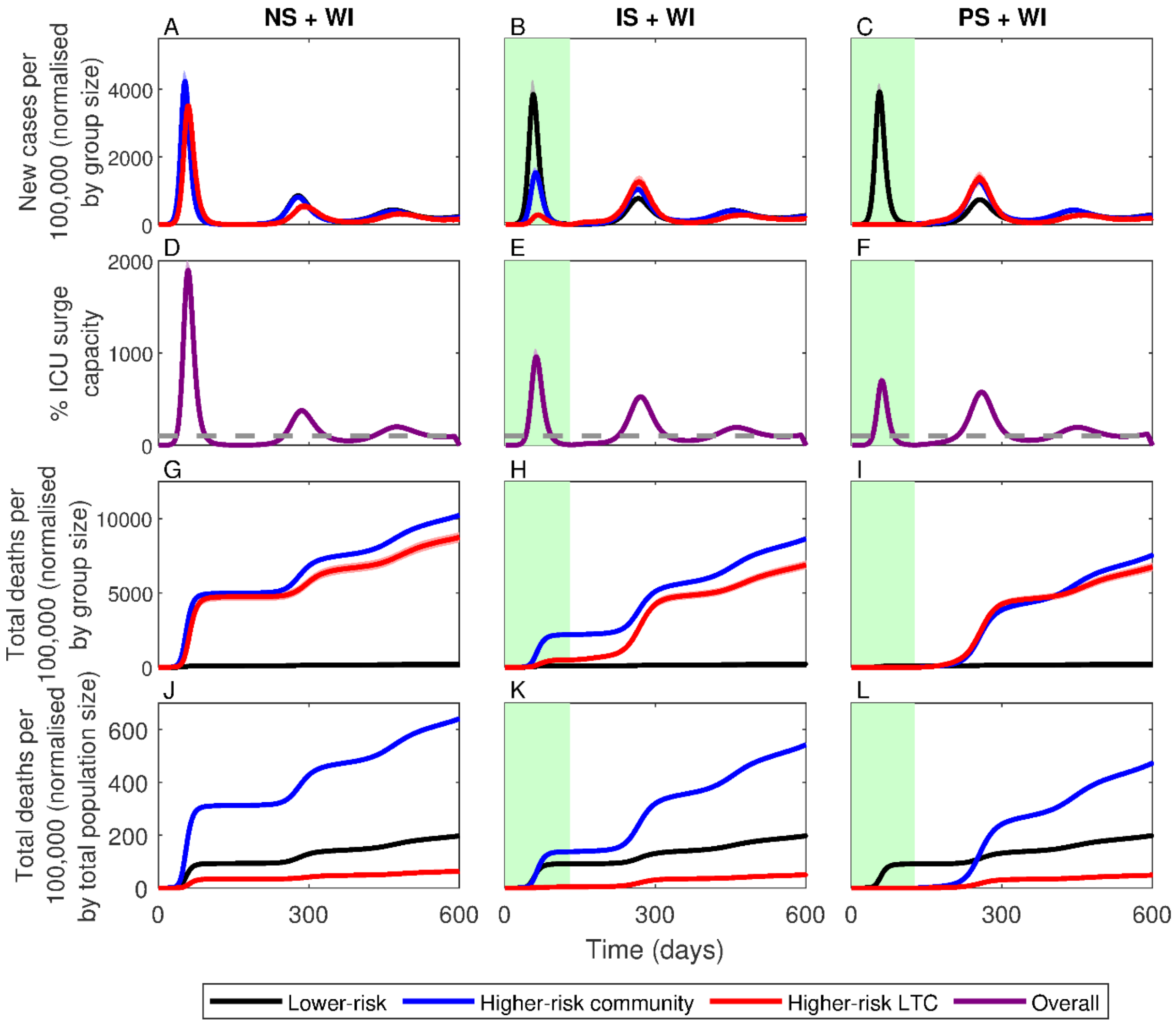

**Fig. 6. Simulations of the three shielding scenarios with waning immunity (+WI).** All other descriptions as in Fig. 2. Lines correspond to means for groups at: lower-risk (black), higher-risk in the community (blue) and in LTC facilities (red), with shading indicating ± 1 SD. Green shading indicates the shielding phase. NS: no shielding; IS: imperfect shielding; PS: perfect shielding.

## §4: Discussion

Our results demonstrate critical epidemiological weaknesses in shielding strategies that aim to achieve herd immunity by isolating the vulnerable while allowing infections to spread among lower-risk members of the population. While our main results focus on a limited set of parameters, our findings are qualitatively robust to sensitivity analysis (§S1 of the *Supplementary material*). Even in the best-case scenario with perfect shielding, our model estimates that there would have been tens of thousands of avoidable deaths among those deemed to be at lower risk due to limited mitigation in this subpopulation, even without accounting for the rapid depletion of healthcare capacity. A significant reduction in contact rates would have been required to avoid overwhelming healthcare capacity during shielding (18), but the population would have then failed to achieve herd immunity, allowing a second, deadlier wave to occur following the lifting of restrictions. Under more realistic assumptions of imperfect shielding, our model estimates that deaths would have been 150% to 300% higher compared to perfect shielding. Breaking down deaths by risk category and location reveals contrasting effects of the scenarios on different groups (Table S2). In some cases (+RC), LTC residents have disproportionately higher death rates than similar individuals in the community, and in others the converse is true (IS). This occurs because LTC residents are clustered together within facilities, whereas higher-risk individuals outside of LTC facilities are assumed to mix randomly in the community. Clustering of susceptible contacts means that higher-risk individuals in LTC facilities are more adversely affected than those in the community when herd immunity is not reached during the shielding phase, as LTC facilities remain vulnerable to large outbreaks once restrictions are lifted. This effect is not seen in previous models which do not differentiate between higher risk members of the population who reside in the community and those who reside in LTC facilities (6).

Our model demonstrates that shielding would have only worked well under practically unrealizable conditions. If any of these conditions had not been met, then significant outbreaks would have occurred in higher-risk subpopulations, leading to many more deaths than if shielding were perfect. To be effective, shielding would have also required those who were at higher risk to not only be rapidly and accurately identified, but also to shield themselves for an indefinite period. If higher-risk individuals were to be misdiagnosed or were unable to fully isolate this would have decreased the effectiveness of shielding. For example, shielding would have been especially difficult for households that contained both higher- and lower-risk individuals (e.g., 74% of CEV people in England live with other people, and 15% live with children aged under 16 years (30)). The large number of multi-risk households suggests that either shielding would have been far from perfect, or a significant proportion of lower-risk individuals would have also had to shield, in which case it would have been harder (or perhaps impossible) to achieve herd immunity during the shielding phase.

The present study focuses on three critical epidemiological weaknesses in shielding strategies, but there are many additional epidemiological, logistical, and ethical problems with shielding that are not captured by our model (9,31). Notably, even if perfect shielding had been possible, there would have been major issues associated with the large number of infections required to achieve herd immunity. Long-term sequalae of infection, known collectively as 'long COVID', are thought to affect between 5 and 10% of those infected (32,33), which would have left many otherwise healthy people with significant long-term health problems. A large epidemic would have also potentially allowed new variants to emerge, which may have been more transmissible, more deadly, or able to escape immunity. We made the conservative assumption of no pathogen evolution, but novel variants would have rendered shielding an even less effective strategy. Our model also made conservative assumptions regarding infection fatality rates (IFRs; see §2.1) and immunity, but more realistic assumptions are likely to make the case for shielding far worse. For example, we used relatively low estimates for the IFRs and assumed that these were fixed even though healthcare capacity would have been significantly overwhelmed under all shielding scenarios. The model also did not capture the impact of healthcare burden on mortality from other causes. We further assumed that immunity from infection was perfect and long-lasting ('best-case' assumptions for shielding), but neither is likely to be true in reality (34). These additional considerations, in combination with the clear flaws indicated by our modelling, suggest that, while an idealized shielding strategy may have allowed populations to achieve herd immunity with fewer deaths, they are likely to have failed catastrophically in practice.

## Supplementary legends

Fig. S1: **Sensitivity analysis – shielding effectiveness.**

Fig. S2: **Sensitivity analysis – altering contact reduction, IS.**

Fig. S3: **Sensitivity analysis – altering contact reduction, PS.**

Fig. S4: **Sensitivity analysis – altering $\boldsymbol{R_0}$, IS.**

Fig. S5: **Sensitivity analysis – altering $\boldsymbol{R_0}$, IS + RC.**

Fig. S6: **Sensitivity analysis – altering intra-LTC contact proportion, IS.**

Fig. S7: **Sensitivity analysis – altering intra-LTC contact proportion, IS + RC.**

Fig. S8: **Sensitivity analysis – altering incubation period, IS.**

Fig. S9: **Sensitivity analysis – altering incubation period, PS.**

Fig. S10: **Sensitivity analysis – altering infectious period, IS.**

Fig. S11: **Sensitivity analysis – altering infectious period, PS.**

Fig. S12: **Sensitivity analysis – altering external infection rate, IS.**

Fig. S13: **Sensitivity analysis – altering external infection rate, PS.**

Fig. S14: **Sensitivity analysis – one LTC size.**

Fig. S15: **Splitting the infectious class into symptomatic and asymptomatic.**

Fig. S16: **Effective reproductive numbers of main text scenarios.**

Table S1: **Shielding parameter values for main text scenarios.**

Table S2: **Death counts for selected main text scenarios.**

Table S3: **Hospitalisation data.**

**Acknowledgments:**

We thank B. Adams, A. Best, G. Constable, E. Feil, T. Rogers, and R. Thompson for helpful discussions and comments on the manuscript.

**Funding:**

BA is funded by Natural Environment Research Council grants NE/N014979/1 and NE/V003909/1. CAS is funded by Natural Environment Research Council grant NE/V003909/1. The funders had no role in study design, data collection and analysis, decision to publish, or preparation of the manuscript.

**Author contributions:**

Conceptualization: CAY, BA

Formal analysis: CAS

Methodology: CAS, CAY, BA

Investigation: CAS

Visualization: CAS

Writing – original draft: CAS, CAY, BA

Writing – review & editing: CAS, CAY, BA

**Competing interests:**

The authors have no competing interests to declare.

**Additional information:**

Supplementary information is available for this paper.

Correspondence and requests for materials should be addressed to Ben Ashby.

**Supplementary Material: Critical weaknesses in shielding strategies for COVID-19**

**Authors:** Cameron A. Smith[1], Christian A. Yates[1]†, Ben Ashby[1,2]†*

**Affiliations:**

[1]Department of Mathematical Sciences, University of Bath; Bath, BA2 7AY, UK.

[2]Milner Centre for Evolution, University of Bath; Bath, BA2 7AY, UK.

*Corresponding author. Email: bna24@bath.ac.uk.

†Equal contribution

**§S1: Sensitivity analysis**

In this section, we present additional results which demonstrate that the qualitative results presented in the main text are robust to reasonable variations in parameters. Specifically, we vary the effectiveness of shielding (Fig. S7) and investigate the effect of changing the reduction in contact by lower-risk individuals during the shielding phase (Fig. S2 and Fig. S3), the basic reproductive number (Fig. S4 and Fig. S5) and the proportion of contacts within LTC facilities (Fig. S6 and Fig. S7). Furthermore, we verify that our results are robust when varying the incubation period (Fig. S8 and Fig. S9), the infectious period (Fig. S10 and Fig. S11), the force of external infection (Fig. S12 and Fig. S13), and robust to variation in LTC facility size (Fig. S14).

We begin by varying the effectiveness of shielding in the imperfect shielding scenario by 10% either side of the baseline of 80% employed in the main text (Fig. S7). Unsurprisingly, decreasing the effectiveness of the shielding increases the number of deaths in the higher-risk subpopulations, while increasing the effectiveness of shielding has the opposite effect. The qualitative epidemiological dynamics results are broadly similar, however.

We next vary the contact reduction among the lower-risk population, from the 50% employed in the main text, to a 40% and 60% reduction in both the perfect shielding (Fig. S) and imperfect shielding (Fig. S3) cases. In the perfect shielding case, a greater reduction in the contact rate among lower-risk members of the population during shielding is associated with a longer shielding phase (420 days on average for 60% reduction, 300 days on average for 50% reduction, 200 days on average for 40% reduction) and a larger second wave, particularly in the higher-risk in the community. We see the same broad pattern when considering the imperfect shielding scenario (Fig. S3).

Next, we investigate the effects of varying the basic reproduction number from $R_0 = 3$ (employed in the main text scenarios) to $R_0 = 2.5$ or $R_0 = 3.5$. We conduct these simulations on both the imperfect shielding (IS; Fig. S) and imperfect shielding with reduced contact (IS + RC; Fig. S) scenarios, as there is very little effect on the results for perfect shielding. For the IS scenario (Fig. S), we see an increase in deaths and infections as the basic reproduction number increases, however the qualitative behaviour remains unchanged. When reduced contact among lower-risk individuals is included (Fig. S), we see that increasing $R_0$ results in a larger and shorter first wave, and a smaller second wave. When $R_0$ is decreased, the first wave is much longer with very few cases, but a notable increase in cases during the second wave.

Now we vary the value of $\lambda$, the proportion of contacts that an LTC resident has within their LTC facility. Again, we conduct this analysis on both the imperfect shielding (IS; Fig. S) and imperfect shielding with reduced contact (IS + RC; Fig. S7**Error! Reference source not found.**) scenarios. In both cases there are no qualitative effects on the results.

In the main text, we fix the incubation and infectious periods at $1/\sigma = 2$ and $1/\Gamma = 5$ respectively. In Figs. S8 and S9 we alter the incubation period and in Figs. S10 and S11 we alter the infectious period. In each of these plots, we see no qualitative differences in our results. As expected, the numbers of deaths are unaltered, with only the infection profiles changing to become either narrower yet taller, or wider and shorter. Similarly, we see no qualitative change in our results when varying the external force of infection with either imperfect shielding (Fig. S12) or perfect shielding (Fig. S13). We also assess whether the distribution of care home sizes affects the outcome by replacing small, medium and large LTC facilities with 180 identical LTCs of size 40. This yields the same number of LTC residents, only they are now distributed into equally sized homes. This redistribution does not alter our conclusions (Fig. S14).

**§S2: Splitting the infection class into symptomatic and asymptomatic classes**

In the main text, in order to ensure that the model remains as simple as possible, we consider just a single infectious class, which combines symptomatic and asymptomatic individuals . Our transmission and IFR parameters reflect this choice and represent ensemble averages over symptomatic and asymptomatic infections. In this section we demonstrate that explicitly separating the infection class into symptomatic and asymptomatic classes does not qualitatively affect our conclusions.

We define $I_{S,i}(t)$ and $I_{A,i}(t)$ to be the number of symptomatic and asymptomatic individuals in the $i^{\text{th}}$ subpopulation at time $t$. We assume that a proportion $\kappa$ of infected individuals become symptomatic, only those who are symptomatic die from the disease, and that symptomatic and asymptomatic individuals may differ in their transmissibility and IFRs (indicated by tildes). Then we can write analogous transition probabilities to those in the main text (equations (7)) with a new state variable for the $i^{\text{th}}$ subpopulation at time $t$, $\underline{\tilde{Y}_i}^t = (S_i, E_i, I_{S,i}, I_{A,i}, R_i D_i)$:

$$\mathbb{P}\left(\underline{\tilde{Y}_i}^{t+\delta t} - \underline{\tilde{Y}_i}^t = (-1,1,0,0,0,0)\right) = \left(\eta_i(t)S_i + \sum_j \frac{Q_{ij}S_i}{N_j}\left(\tilde{\beta}_{ij}^S I_{S,j} + \tilde{\beta}_{ij}^A I_{A,j}\right)\right)\delta t,$$

$$\mathbb{P}\left(\underline{\tilde{Y}_i}^{t+\delta t} - \underline{\tilde{Y}_i}^t = (0,-1,1,0,0,0)\right) = \sigma\kappa E_i \delta t,$$

$$\mathbb{P}\left(\underline{\tilde{Y}_i}^{t+\delta t} - \underline{\tilde{Y}_i}^t = (0,-1,0,1,0,0)\right) = \sigma(1-\kappa) E_i \delta t,$$

$$\mathbb{P}\left(\underline{\tilde{Y}_i}^{t+\delta t} - \underline{\tilde{Y}_i}^t = (0,0,-1,0,1,0)\right) = \Gamma(1-\tilde{\alpha}_i) I_{S,i} \delta t,$$

$$\mathbb{P}\left(\underline{\tilde{Y}_i}^{t+\delta t} - \underline{\tilde{Y}_i}^t = (0,0,-1,0,0,1)\right) = \Gamma\tilde{\alpha}_i I_{S,i} \delta t,$$

$$\mathbb{P}\left(\underline{\tilde{Y}_i}^{t+\delta t} - \underline{\tilde{Y}_i}^t = (0,0,0,-1,1,0)\right) = \Gamma I_{A,i} \delta t.$$

The transmission rates $\tilde{\beta}_{ij}^S$ and $\tilde{\beta}_{ij}^A$ are similar, but with the symptomatic case being larger than the asymptomatic case. Note that the transmission parameter is constructed from the probability of transmission per contact, the average number of contacts a person has per unit time, the contact matrix $p_{ij}$ and the shielding matrix $Q_{ij}$. We assume that in both symptomatic and asymptomatic cases, the contact and shielding matrices remain the same as these are functions of the subpopulation to which an individual belongs. We assume that symptomatic individuals shed more virus and so have a higher probability of onward transmission per contact, but have a lower average number of contacts per unit time as they are more likely to self-isolate due to symptoms. We run simulations for each of the shielding scenarios with no modifications, setting $\tilde{\beta}_{ij}^{S} = 1.1\beta_{ij}$ and $\tilde{\beta}_{ij}^{A} = 0.8\beta_{ij}$, where $\beta_{ij}$ is the same transmission rate as in the main text. Further, we set $\kappa = 2/3$, and $\tilde{\alpha}_i = \alpha_i/\kappa$ for symptomatic individuals, with $\tilde{\alpha}_i = 0$ for asymptomatic individuals. The results from this can be observed in Fig. S15 and show no qualitative difference to those in the main text Fig. 2.

**§S3: The effective reproductive number**

In order to estimate the effective reproductive number, $R_e(t)$, we multiply the basic reproductive number $R_0$ calculated as in §2.2 in the main text by the proportion of susceptible individuals at time $t$:

$$R_e(t) = \frac{R_0(t)S(t)}{N}$$

Note that we have explicitly written $R_0$ as a function of time here because it depends on whether we are in the shielding period. If we are within the shielding period, the value corresponds to the largest root of the characteristic equation (16a) with the appropriate shielding parameter values. When we are outside of the shielding period, the value of $R_0$ is $\beta_0 r/\Gamma$. We show these plots in time for each of the main text scenarios in Fig. S16. As expected, reducing the contact of lower-risk individuals during shielding results in the effective reproductive number increasing above one when shielding ends. Waning immunity also sees the effective reproductive number oscillate around 1 as the pool of susceptibles shrinks and grows with the waxing and waning of each epidemic wave. In the other two cases (no modifiers and external infection), the overall effective reproductive number does not exceed one once shielding ends. Again, this is to be expected as we have reached herd immunity, but this only confers indirect protection, so those in the lesser exposed subpopulations are still vulnerable to local outbreaks.

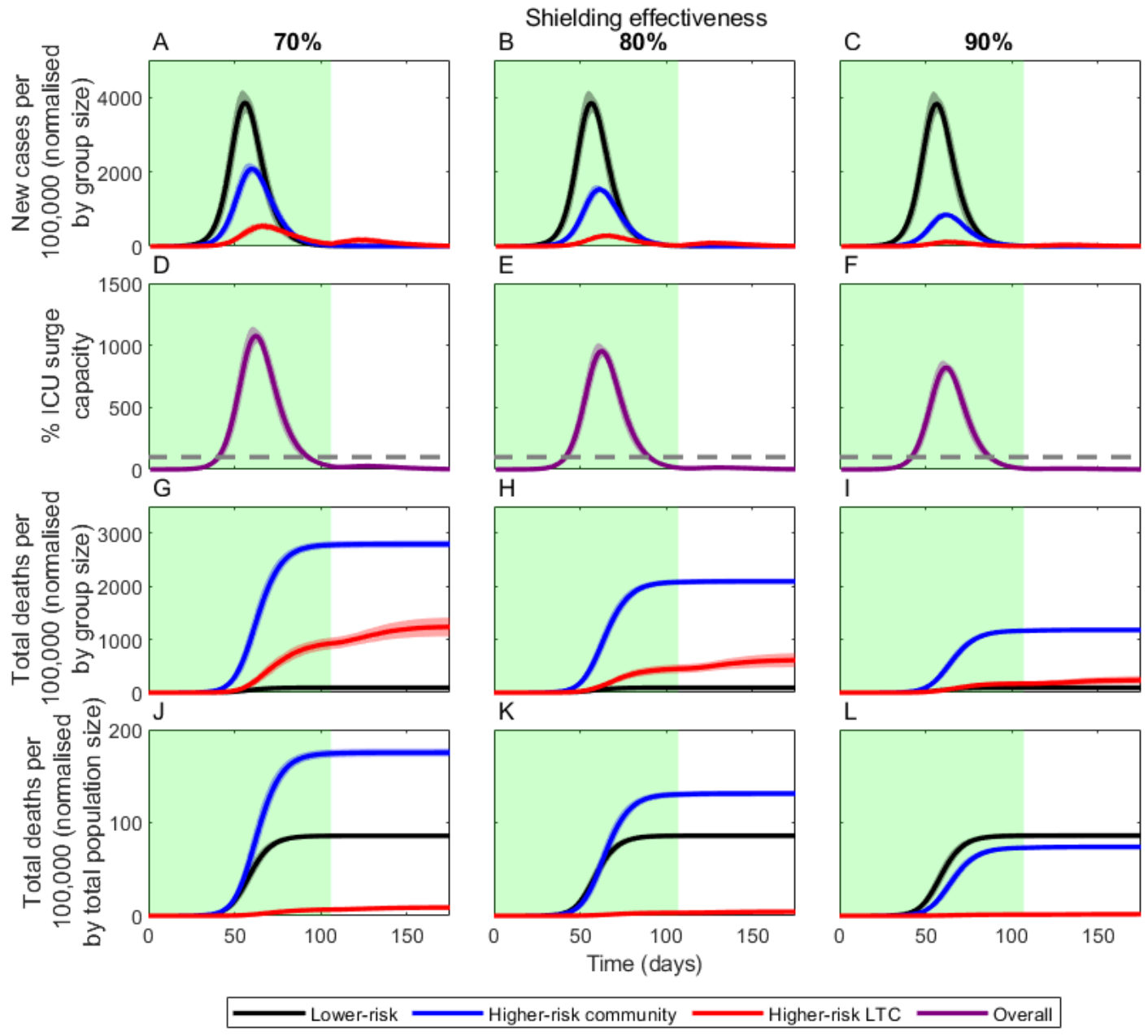

Fig. S7: Effectiveness of imperfect shielding. Baseline figure (reproduced from Fig. 1 in the main text) is in the central column (80% effective), with a 10% difference on either side (70% for column 1 and 90% for column 3). All colours and descriptions are the same as Fig. 2 of the main text. *Lines correspond to means for groups at: lower-risk (black), higher-risk in the community (blue) and in LTC facilities (red), with shading indicating ± 1 SD. Green shading indicates the shielding phase.*

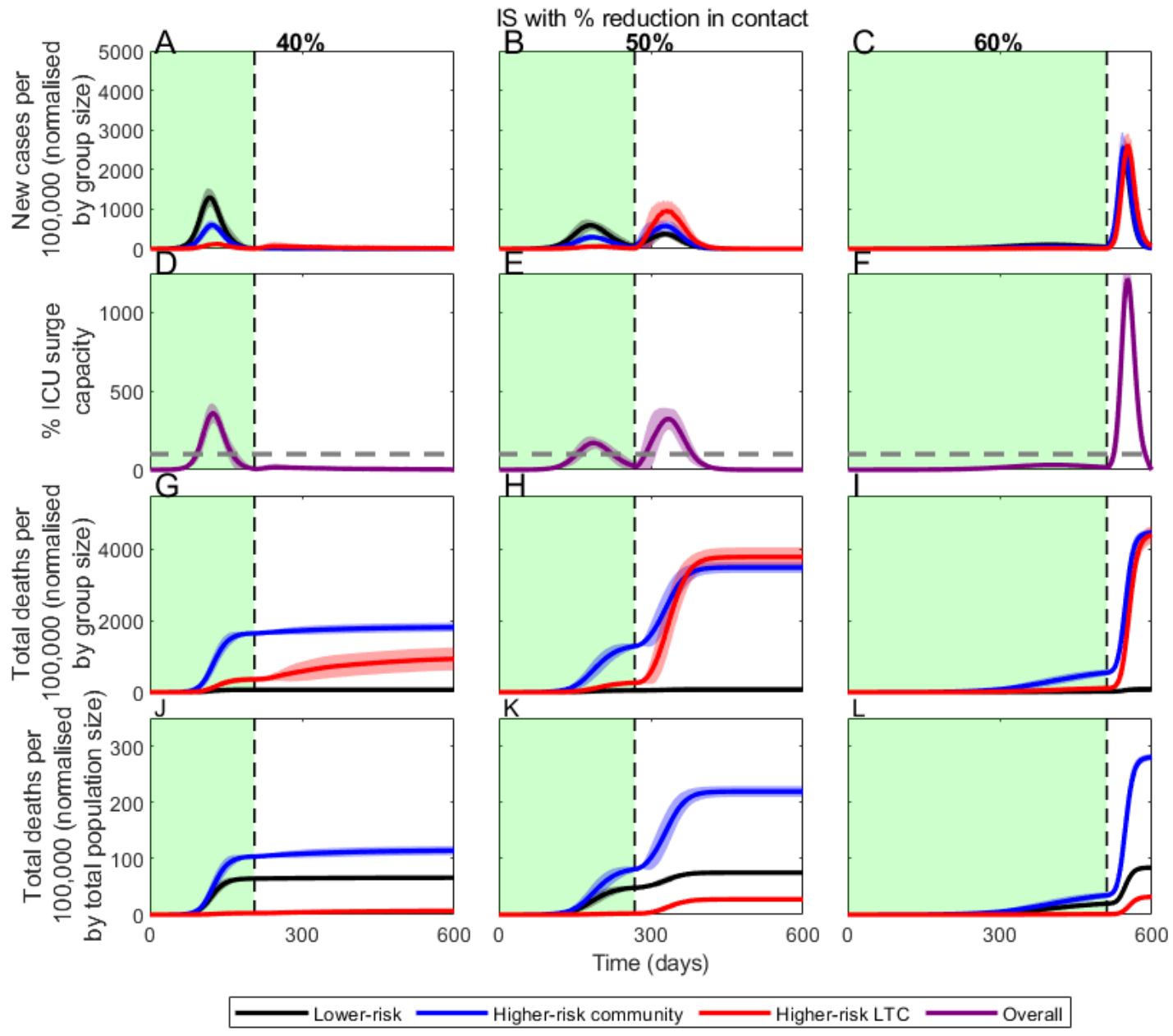

*Fig. S2: The effects of altering the reduction in contacts among lower-risk individuals during the shielding phase under imperfect shielding. The case from the main text (50% reduction, reproduced from Fig. 3) is shown in the second column, with a change of -/+ 10% either side in columns 1 and 3 respectively. All colours and descriptions are the same as Fig. 2 of the main text. Lines correspond to means for groups at: lower-risk (black), higher-risk in the community (blue) and in LTC facilities (red), with shading indicating ± 1 SD. Green shading indicates the shielding phase.*

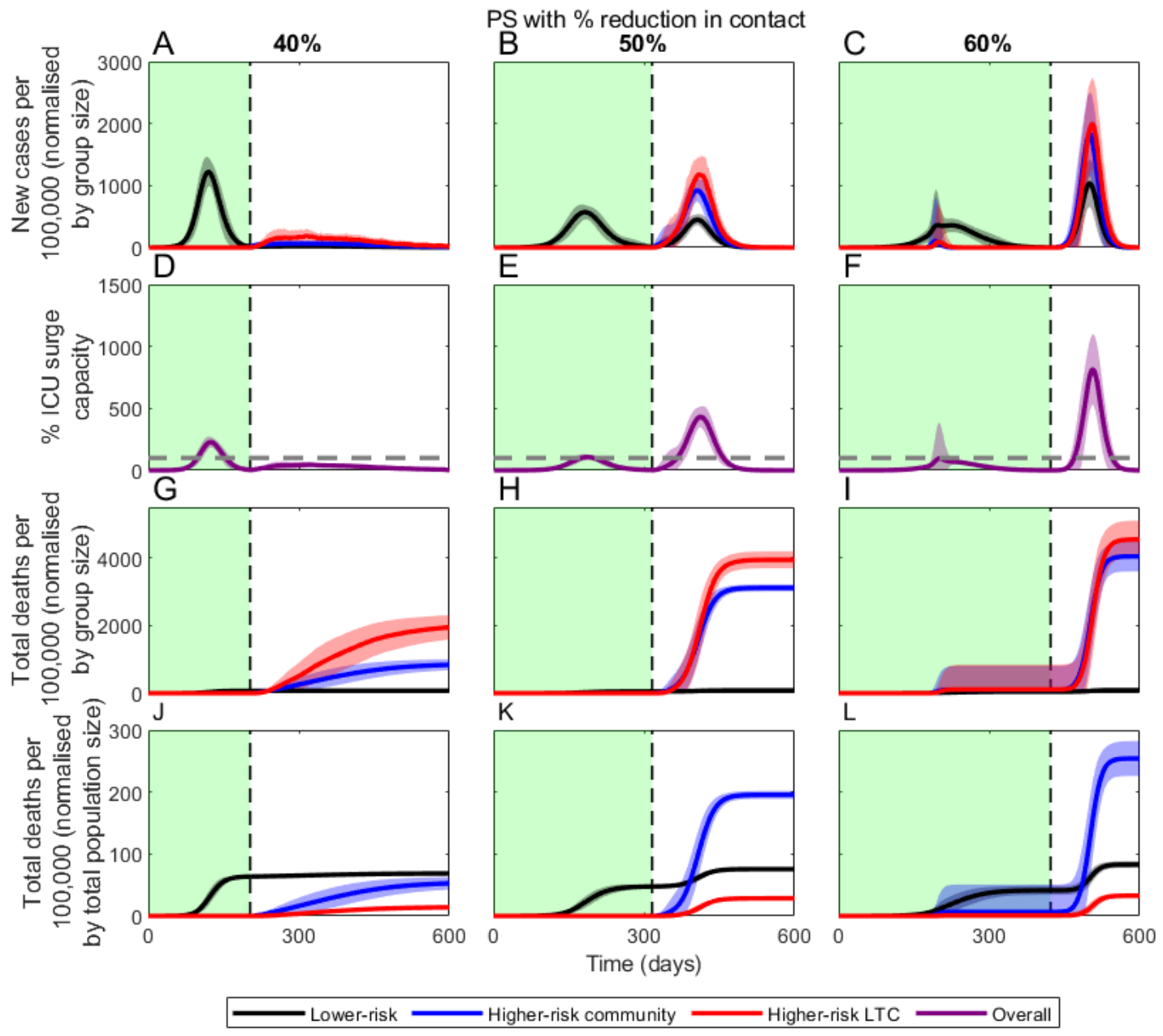

*Fig. S3: The effects of altering the reduction in contacts among lower-risk individuals during the shielding phase under perfect shielding. The case from the main text (50% reduction, reproduced from Fig. 3) is shown in the second column, with a change of -/+ 10% either side in columns 1 and 3 respectively. All colours and descriptions are the same as Fig. 2 of the main text. Lines correspond to means for groups at: lower-risk (black), higher-risk in the community (blue) and in LTC facilities (red), with shading indicating ± 1 SD. Green shading indicates the shielding phase.*

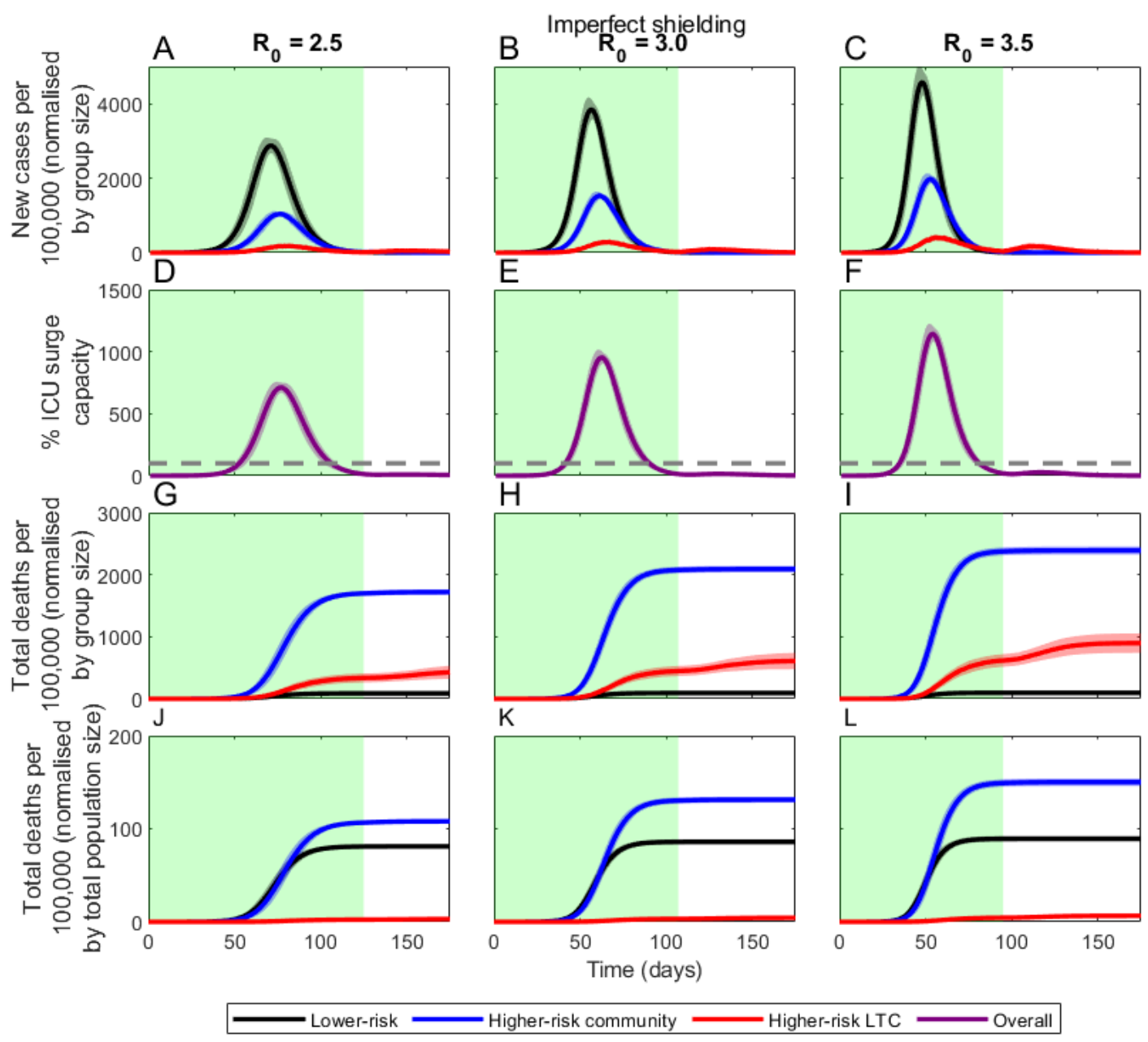

*Fig. S4: Altering $R_0$ under the imperfect shielding scenario. The second column is reproduced from Fig. 1. Column 1 represents a reduction in $R_0$ to 2.5 and column 3 an increase to 3.5. All colours and descriptions are the same as Fig. 2 of the main text. Lines correspond to means for groups at: lower-risk (black), higher-risk in the community (blue) and in LTC facilities (red), with shading indicating ± 1 SD. Green shading indicates the shielding phase.*

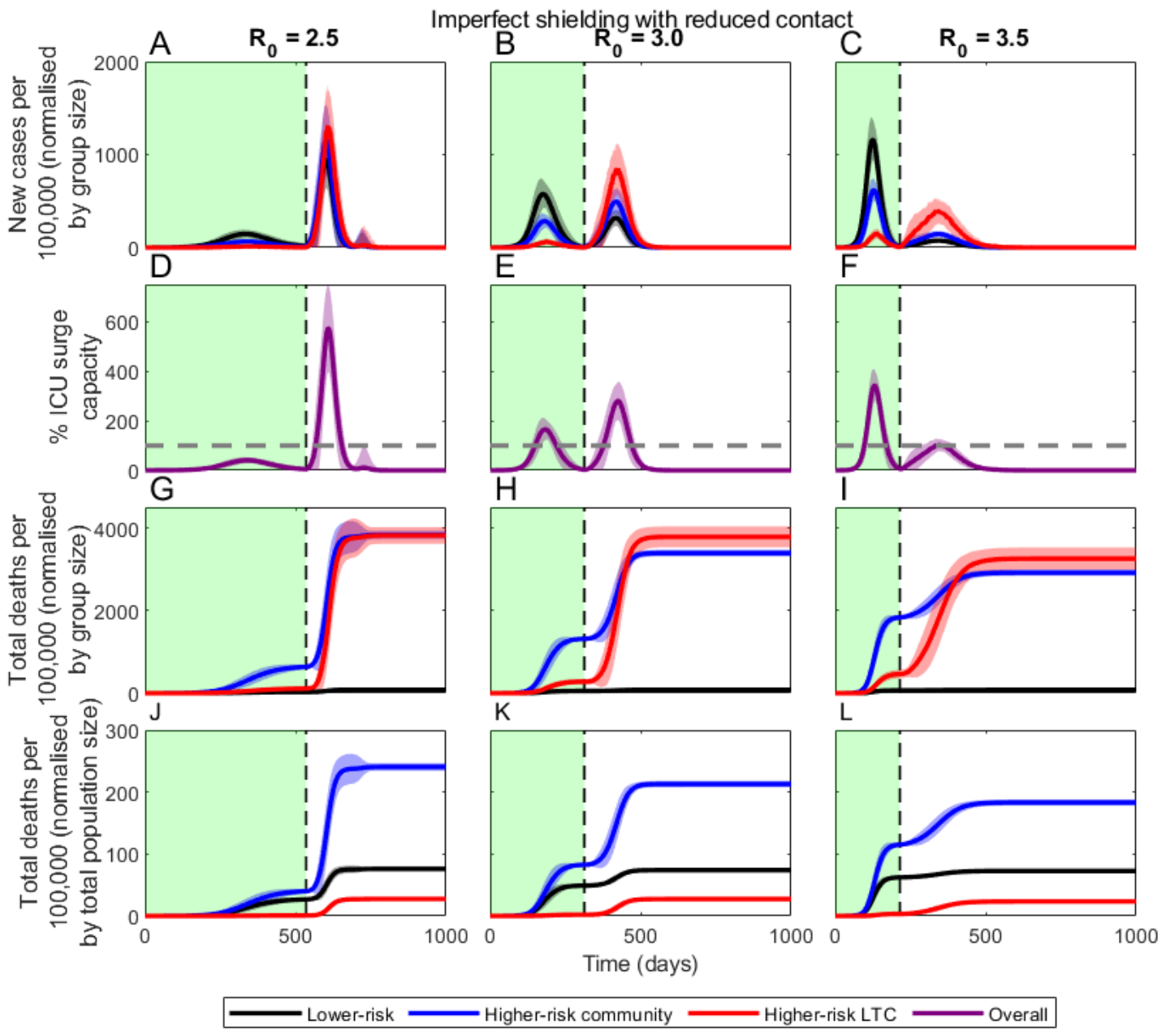

*Fig. S5: Altering $R_0$ under the imperfect shielding scenario with reduced contact. The second column is reproduced from Fig. 4. Column 1 represents a reduction in* $R_0$ *to 2.5 and column 3 an increase to 3.5. All colours and descriptions are the same as Fig. 2 of the main text. Lines correspond to means for groups at: lower-risk (black), higher-risk in the community (blue) and in LTC facilities (red), with shading indicating ± 1 SD. Green shading indicates the shielding phase. NS: no shielding; IS: imperfect shielding; PS: perfect shielding.*

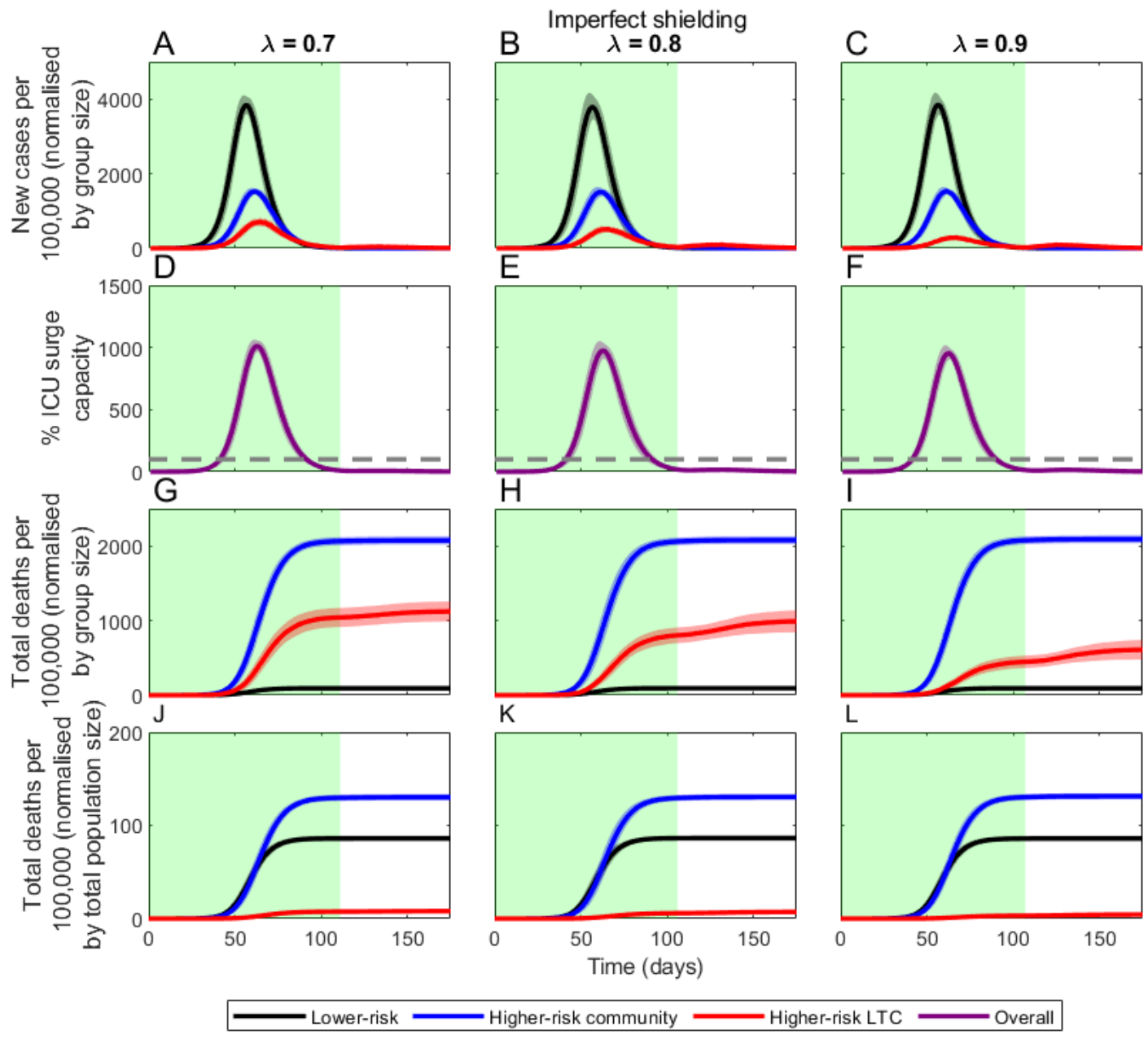

*Fig. S6: Altering $\lambda$ under the imperfect shielding scenario. The third column is reproduced from Fig. 1. Column 1 represents a reduction in $\lambda$ to 0.7 and column 2 an decrease to 0.8. All colours and descriptions are the same as Fig. 2 of the main text. Lines correspond to means for groups at: lower-risk (black), higher-risk in the community (blue) and in LTC facilities (red), with shading indicating ± 1 SD. Green shading indicates the shielding phase.*

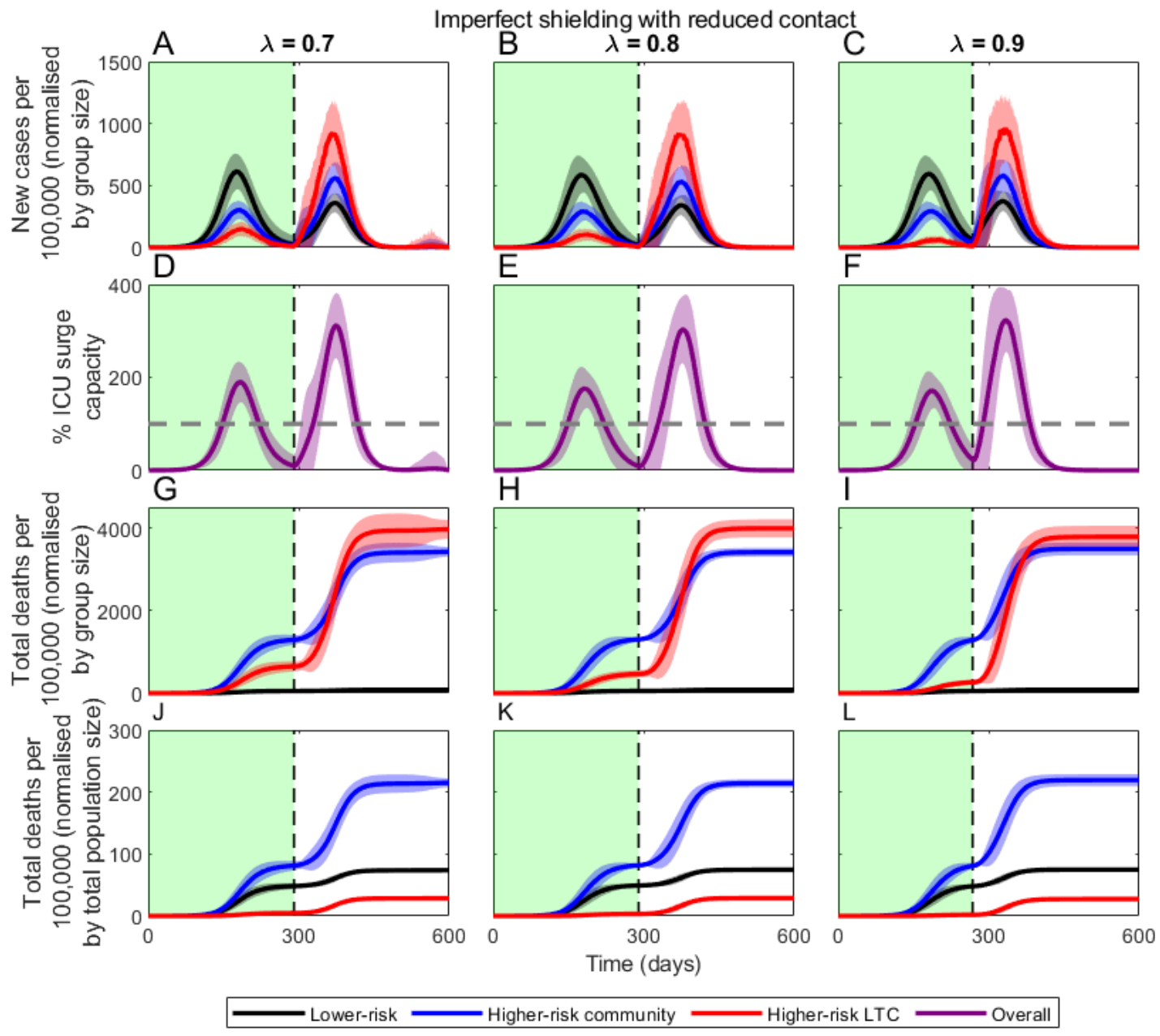

*Fig. S7: Altering $\lambda$ under the imperfect shielding scenario with reduced contact. The third column is reproduced from Fig. 4. Column 1 represents a reduction in $\lambda$ to 0.7 and column 2 an decrease to 0.8. All colours and descriptions are the same as Fig. 2 of the main text. Lines correspond to means for groups at: lower-risk (black), higher-risk in the community (blue) and in LTC facilities (red), with shading indicating ± 1 SD. Green shading indicates the shielding phase.*

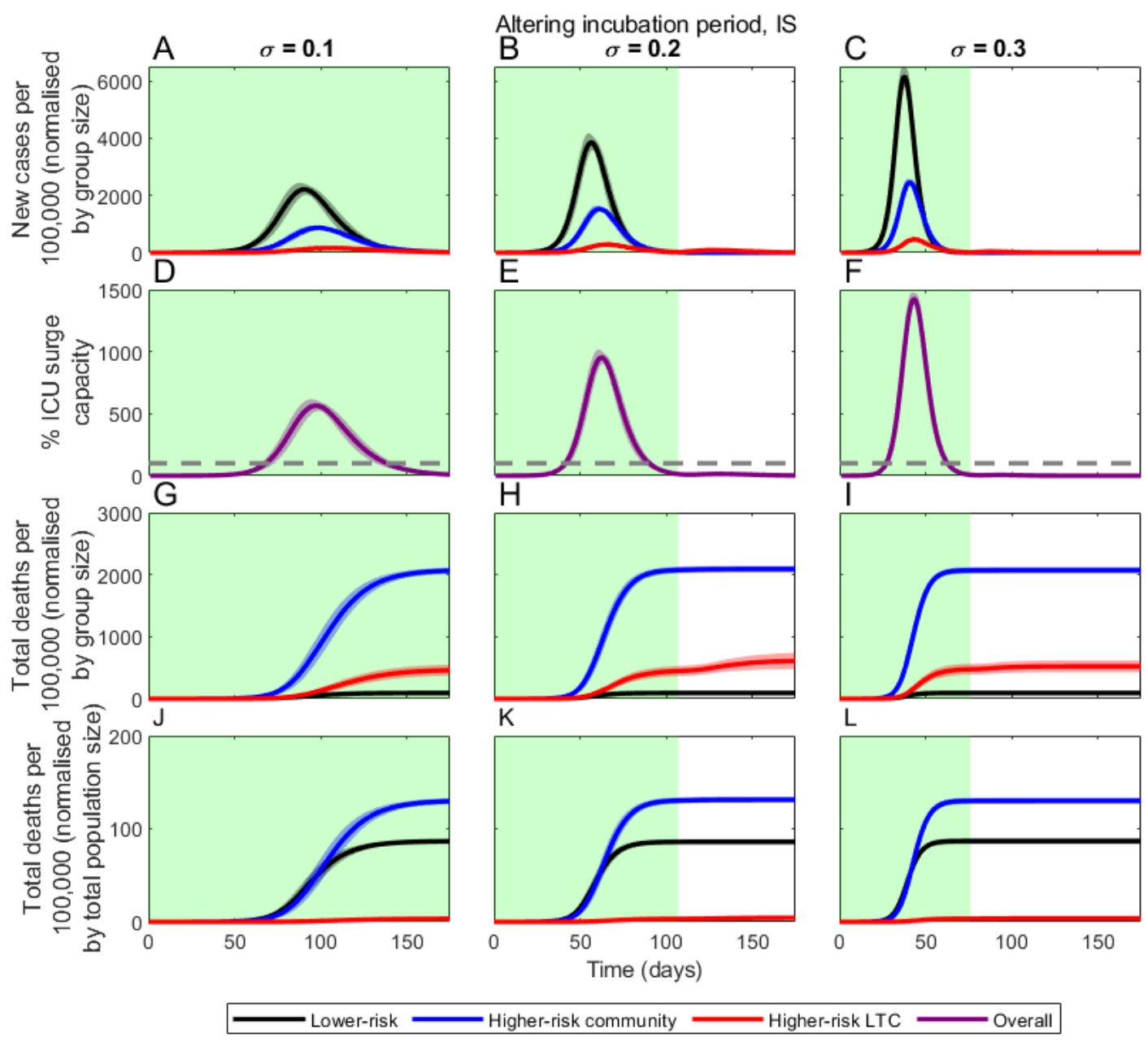

*Fig. S8: Altering the incubation period $1/\sigma$ under the imperfect shielding scenario. The second column is reproduced from Fig. 2. Column 1 represents a reduction in $\sigma$ to 0.1 and column 3 an increase to 0.3. All colours and descriptions are the same as Fig. 2 of the main text. Lines correspond to means for groups at: lower-risk (black), higher-risk in the community (blue) and in LTC facilities (red), with shading indicating ± 1 SD. Green shading indicates the shielding phase.*

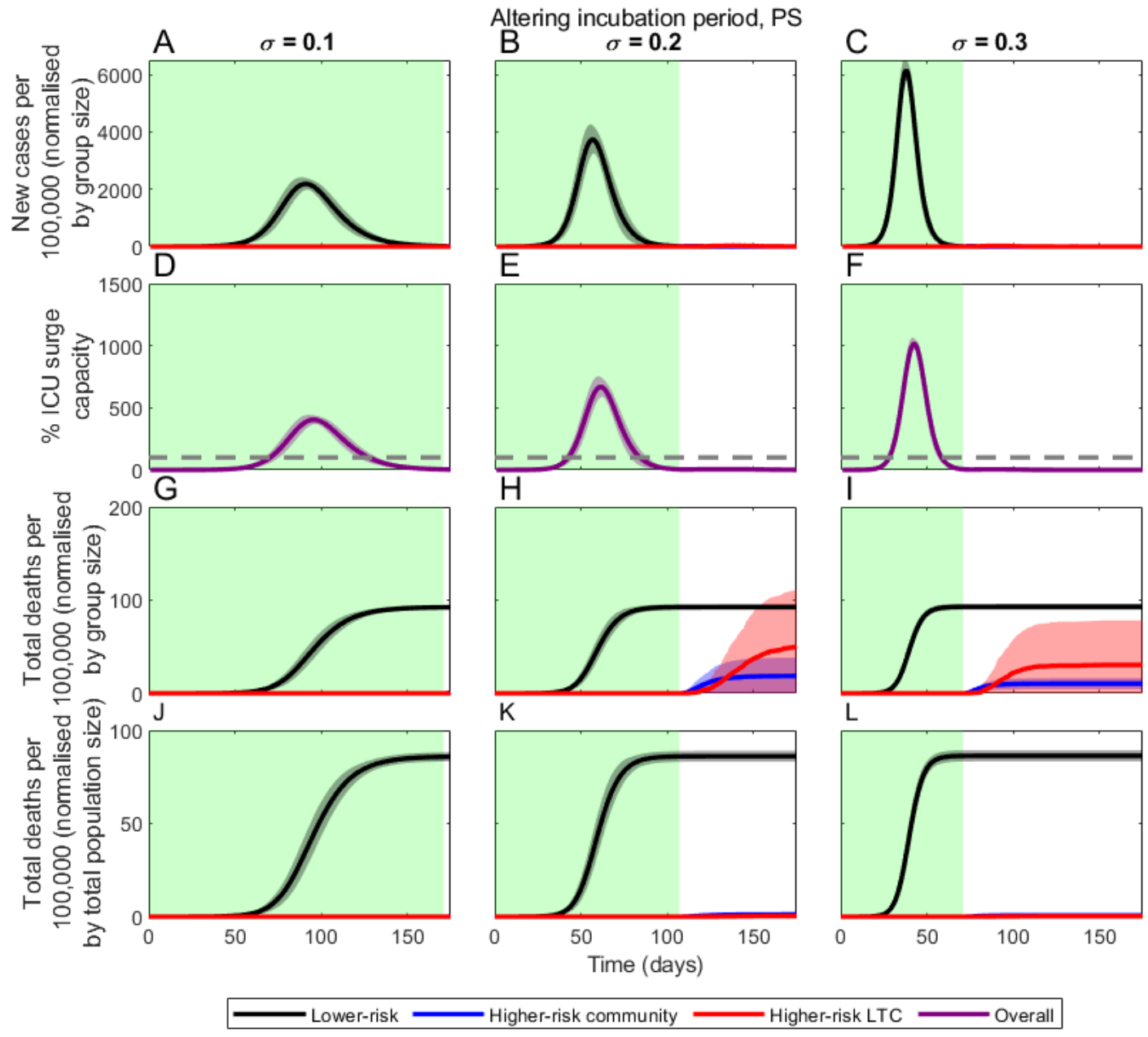

*Fig. S9: Altering the incubation period $1/\sigma$ under the perfect shielding scenario. The second column is reproduced from Fig. 2. Column 1 represents a reduction in $\sigma$ to 0.1 and column 3 an increase to 0.3. All colours and descriptions are the same as Fig. 2 of the main text. Lines correspond to means for groups at: lower-risk (black), higher-risk in the community (blue) and in LTC facilities (red), with shading indicating ± 1 SD. Green shading indicates the shielding phase.*

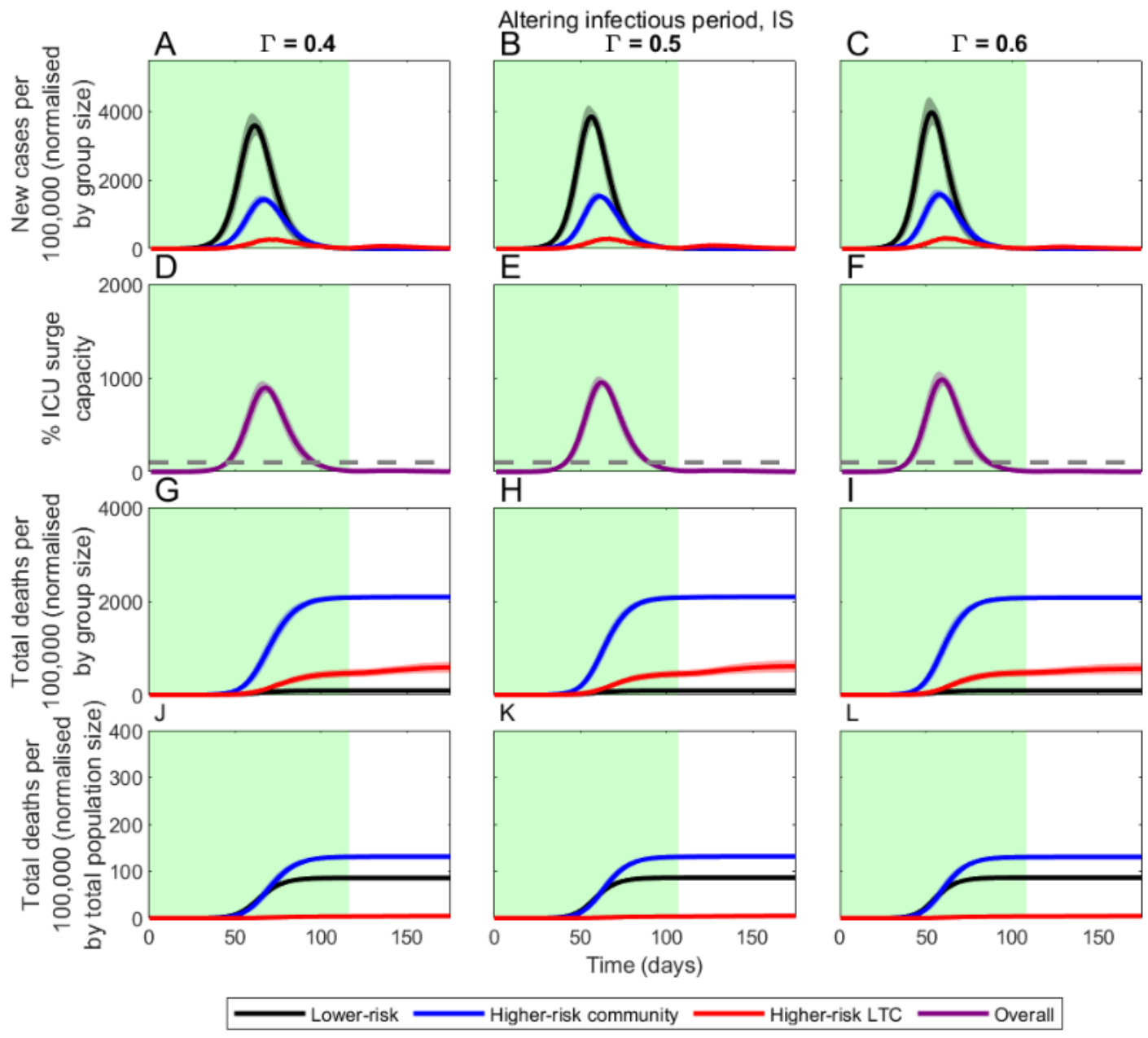

*Fig. S10: Altering the infectious period $1/\Gamma$ under the imperfect shielding scenario. The second column is reproduced from Fig. 2. Column 1 represents a reduction in $\Gamma$ to 0.4 and column 3 an increase to 0.6. All colours and descriptions are the*

*same as Fig. 2 of the main text. Lines correspond to means for groups at: lower-risk (black), higher-risk in the community (blue) and in LTC facilities (red), with shading indicating ± 1 SD. Green shading indicates the shielding phase.*

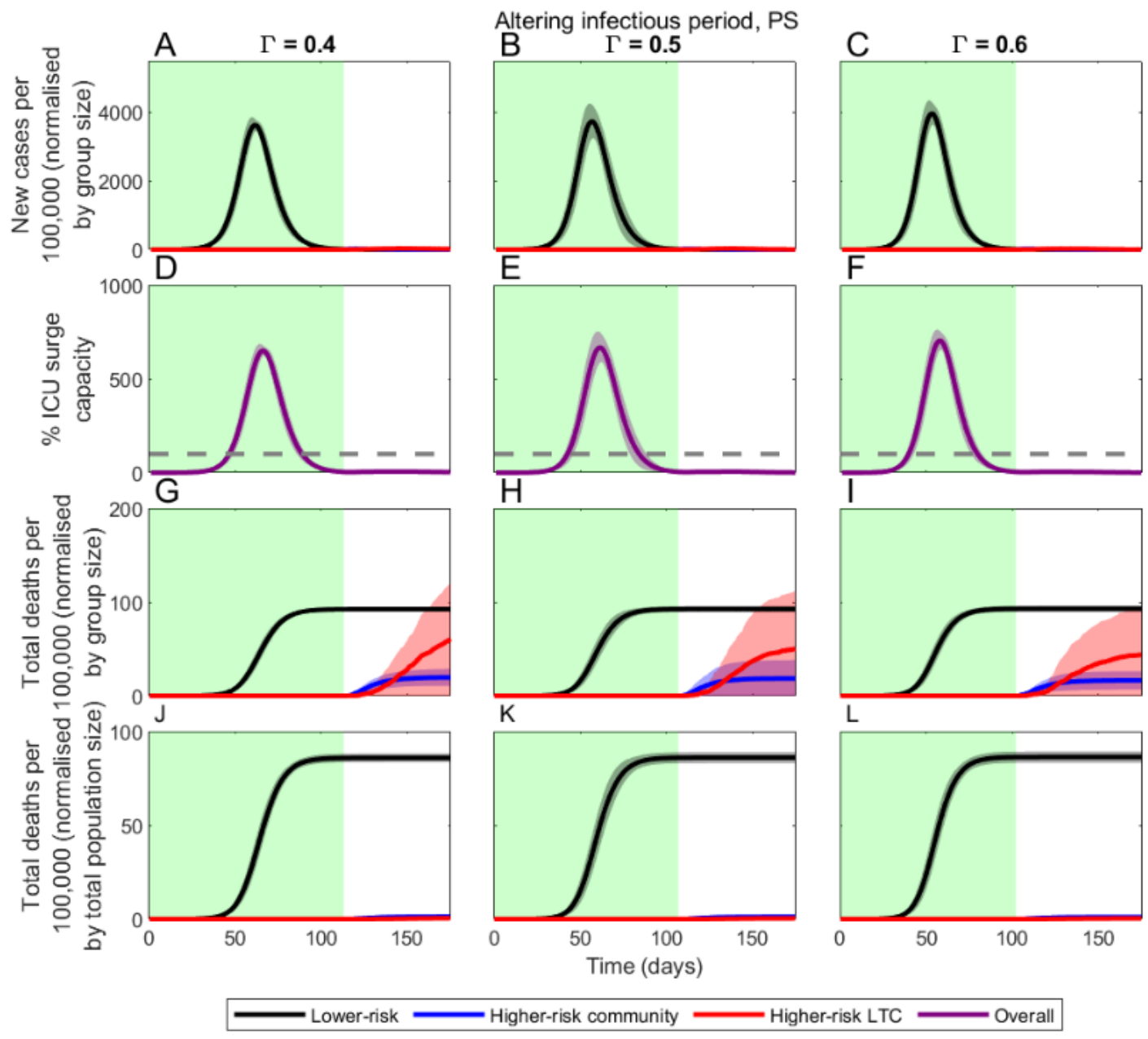

*Fig. S11: Altering the infectious period $1/\Gamma$ under the perfect shielding scenario. The second column is reproduced from Fig. 2. Column 1 represents a reduction in $\Gamma$ to 0.4 and column 3 an increase to 0.6. All colours and descriptions are the same as Fig. 2 of the main text. Lines correspond to means for groups at: lower-risk (black), higher-risk in the community (blue) and in LTC facilities (red), with shading indicating ± 1 SD. Green shading indicates the shielding phase.*

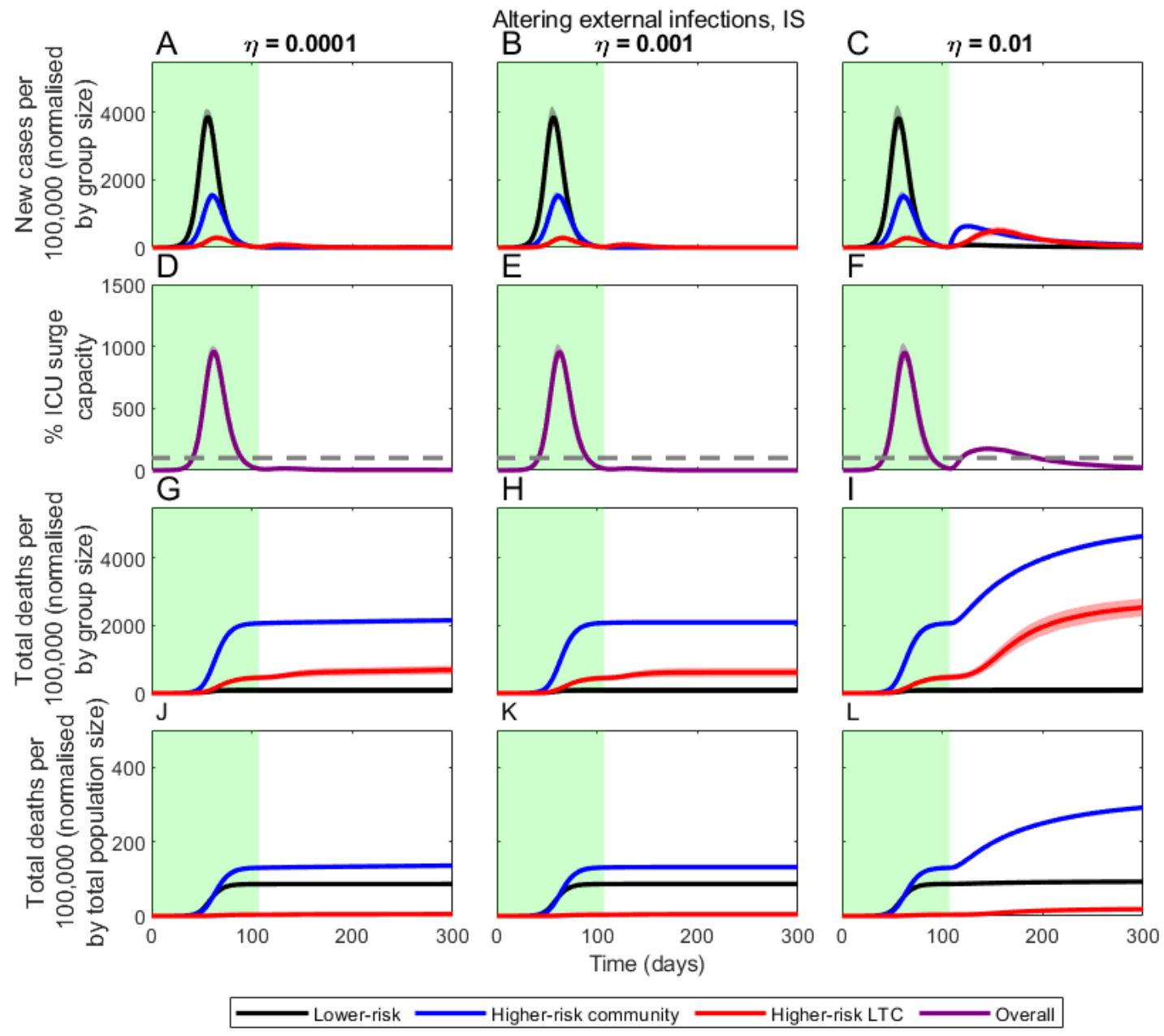

*Fig. S12: Altering the external force of infection, η, under the imperfect shielding scenario. The second column is reproduced from Fig. 5. Column 1 represents a reduction in η to 0.0001 and column 3 an increase to 0.01. All colours and descriptions are the same as Fig. 2 of the main text. Lines correspond to means for groups at: lower-risk (black), higher-risk in the community (blue) and in LTC facilities (red), with shading indicating ± 1 SD. Green shading indicates the shielding phase.*

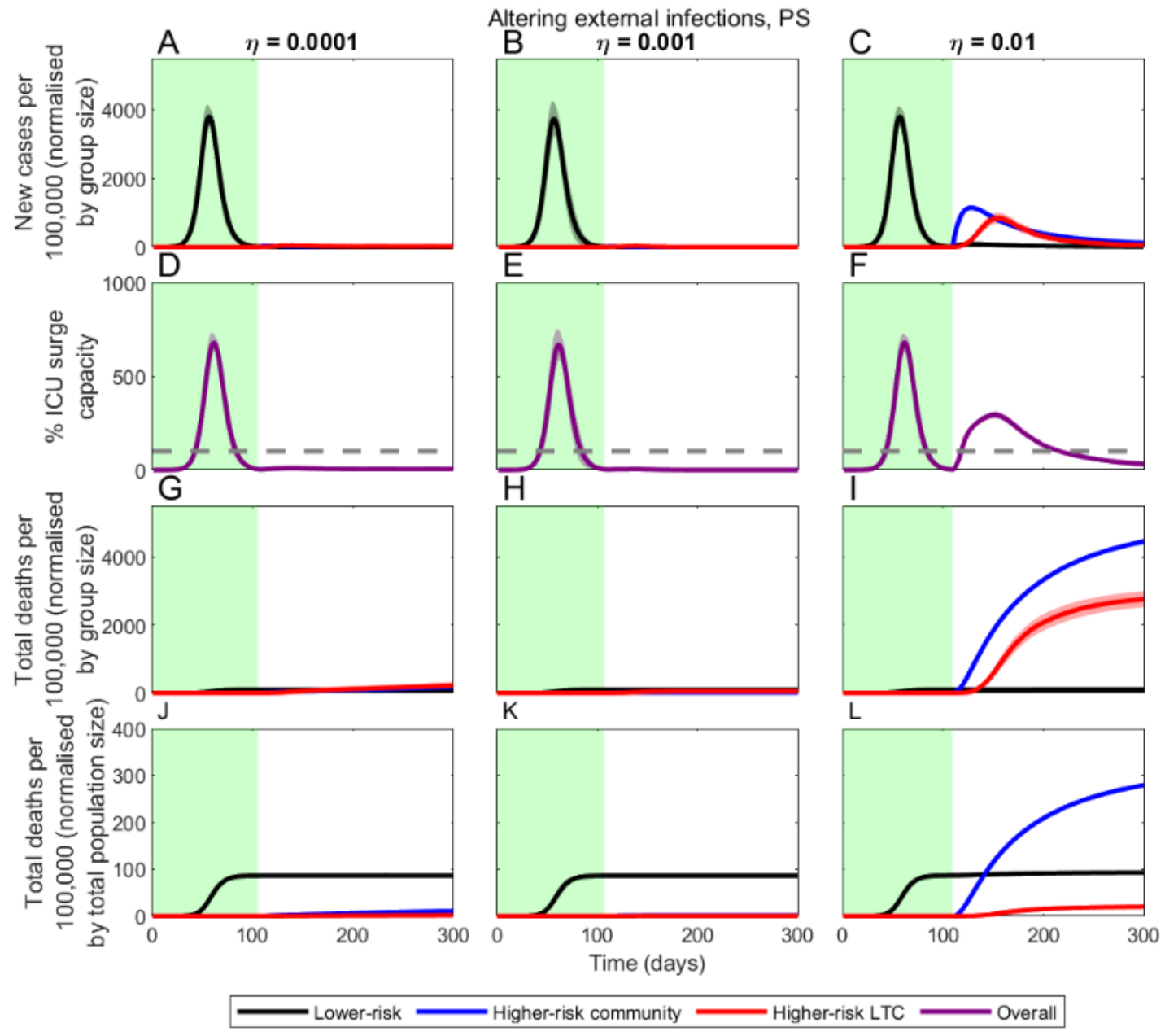

*Fig. S13: Altering the external force of infection, η, under the perfect shielding scenario. The second column is reproduced from Fig. 5. Column 1 represents a reduction in η to 0.0001 and column 3 an increase to 0.01. All colours and descriptions are the same as Fig. 2 of the main text. Lines correspond to means for groups at: lower-risk (black), higher-risk in the community (blue) and in LTC facilities (red), with shading indicating ± 1 SD. Green shading indicates the shielding phase.*

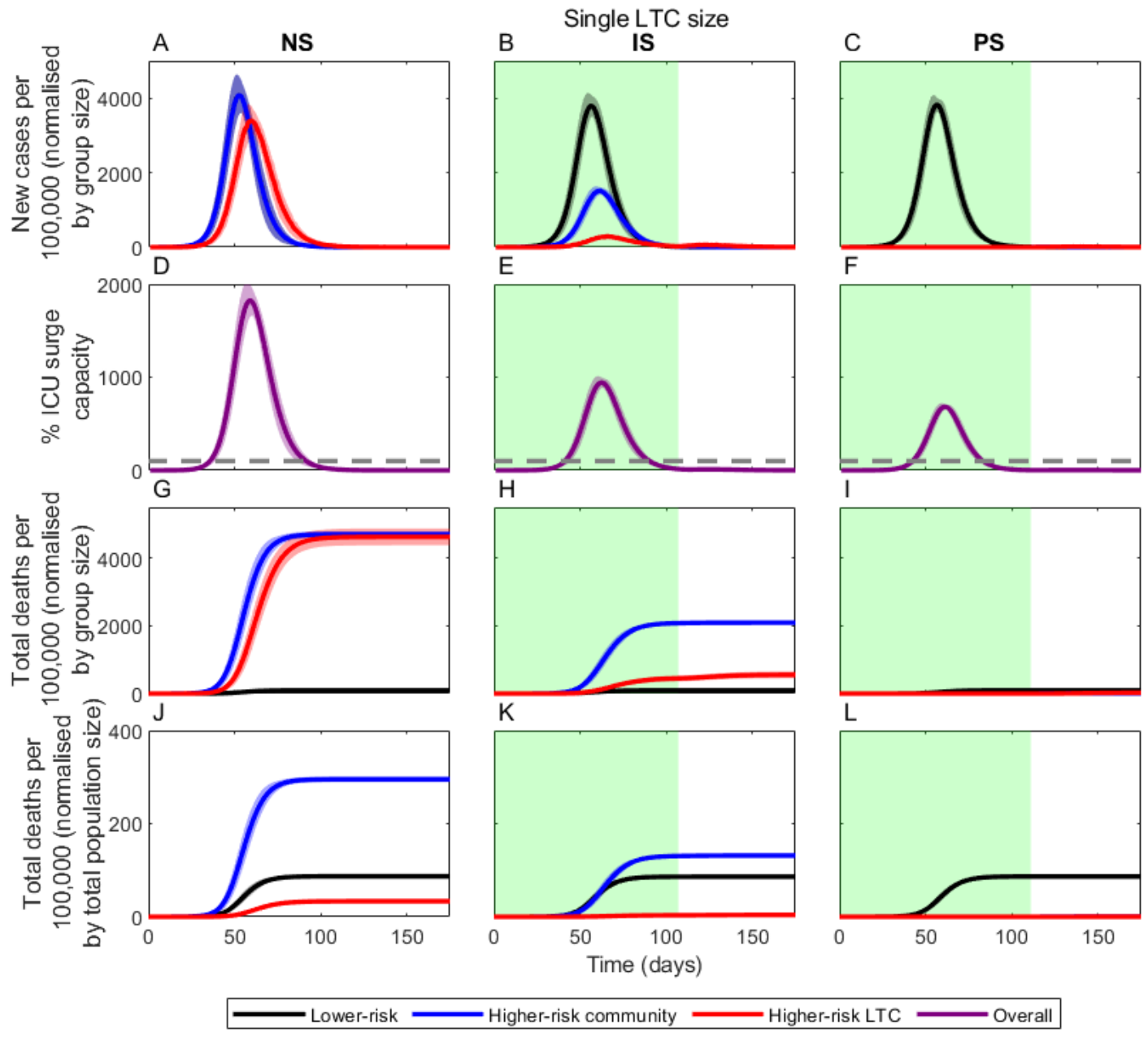

*Fig. S14: Results with a single LTC facility size (180 LTCs each with 40 residents), in contrast to Fig. 2 of the main text where we have small, medium and large LTC facilities. All colours and descriptions are the same as Fig. 2 of the main text. Lines correspond to means for groups at: lower-risk (black), higher-risk in the community (blue) and in LTC facilities (red), with shading indicating ± 1 SD. Green shading indicates the shielding phase. NS: no shielding; IS: imperfect shielding; PS: perfect shielding.*

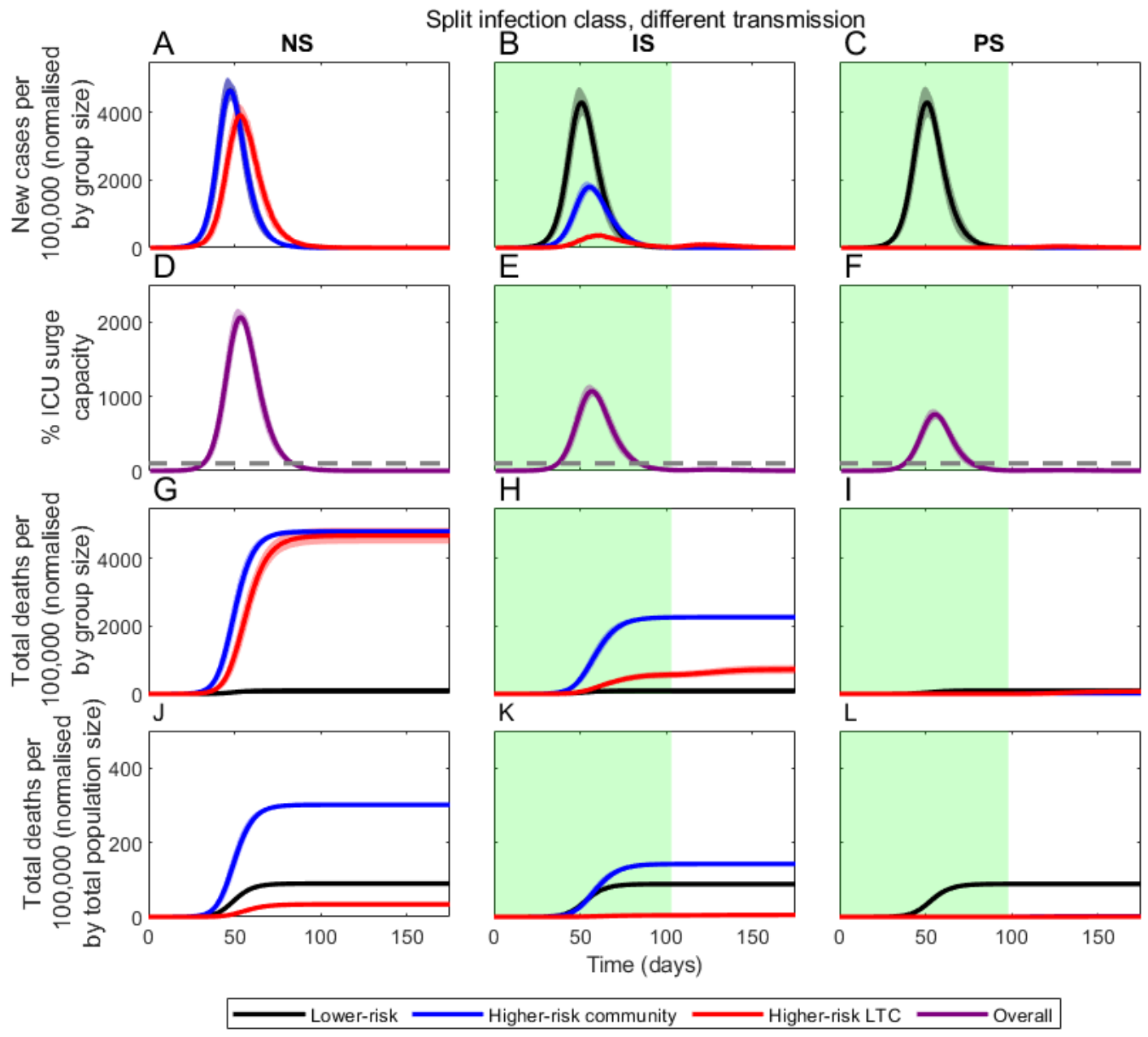

*Fig. S15: Results with separate asymptomatic and symptomatic classes. All colours and descriptions are the same as Fig. 2 of the main text. Lines correspond to means for groups at: lower-risk (black), higher-risk in the community (blue) and in LTC facilities (red), with shading indicating ± 1 SD. Green shading indicates the shielding phase. NS: no shielding; IS: imperfect shielding; PS: perfect shielding.*

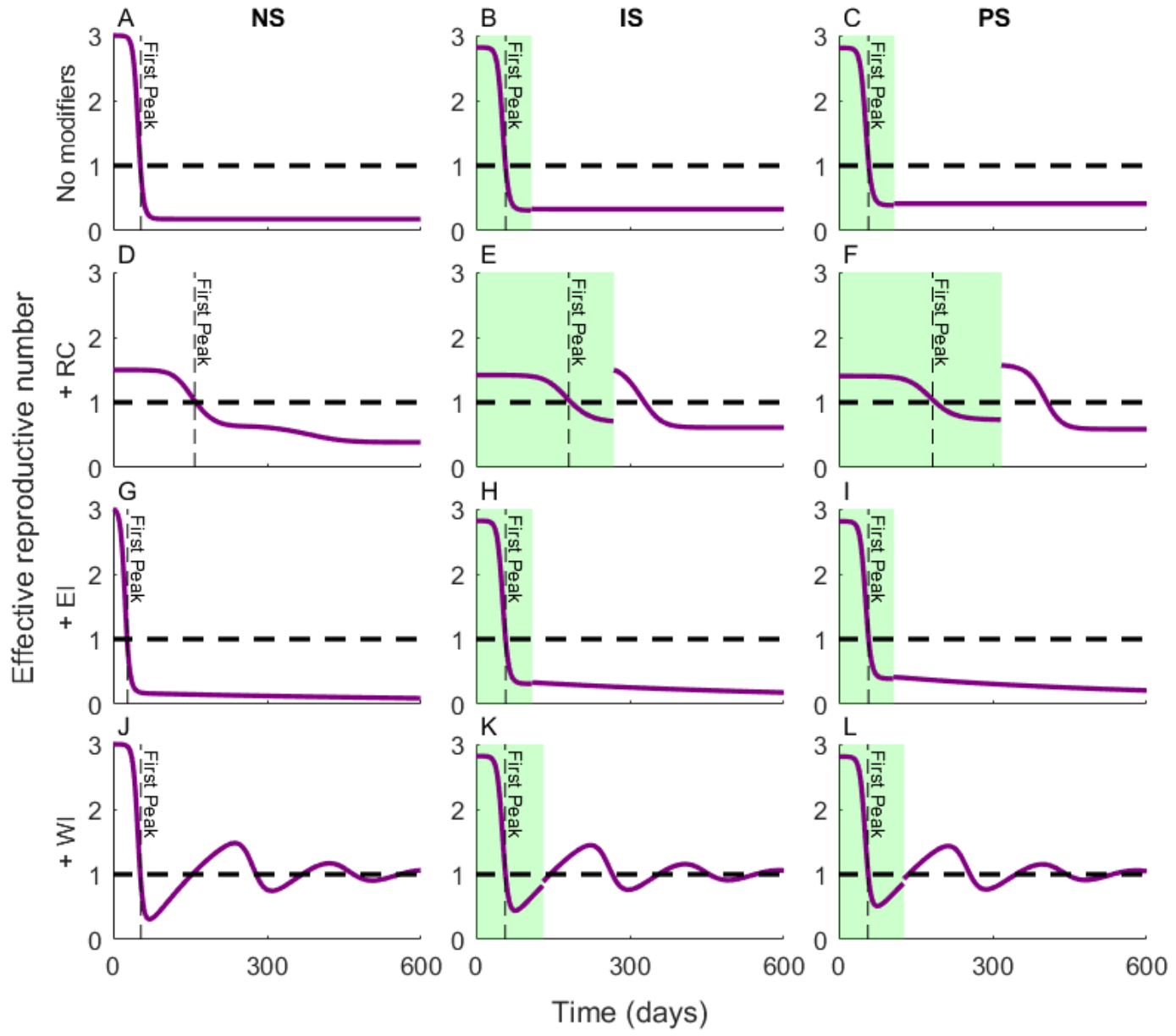

Fig. S16: The effective reproductive number for each of the scenarios described in the main text. The first column represents no shielding (NS), the middle column imperfect shielding (IS) and the third column perfect shielding (PS). Row 1 corresponds to no additional modifiers, row 2 includes reduced contact (RC) at 50%, row 3 adds external infections (EI) after shielding and row 4 has waning immunity (WI). *The black horizontal dashed line denotes the effective reproductive number equalling one, and the grey horizontal line is the time at which the first peak occurs in the incidence curves, coinciding with the point that the effective reproductive number hits 1.*

| Scenario | Shielding parameter | | | | | |
|---|---|---|---|---|---|---|
| | $q_1^s$ | $q_2^S$ | $q_3^S$ | $q_4^S$ | $q_5^S$ | $q_6^s$ |
| NS (+WI) | 1 | 1 | 1 | 1 | 1 | 1 |
| IS (+WI) | 1 | 0.2 | 0.2 | 0.2 | 0.2 | 0 |
| PS (+WI) | 1 | 0 | 0 | 0 | 0 | 0 |
| NS + RC | 0.5 | 0.5 | 0.5 | 0.5 | 0.5 | 0.5 |
| IS + RC | 0.5 | 0.2 | 0.2 | 0.2 | 0.2 | 0 |
| PS + RC | 0.5 | 0 | 0 | 0 | 0 | 0 |
| NS + EI | 1 | 1 | 1 | 1 | 1 | 1 |
| IS + EI | 1 | 0.2 | 0.2 | 0.2 | 0.2 | 0 |
| PS + EI | 1 | 0 | 0 | 0 | 0 | 0 |

*Table S1: Parameter values for the different shielding scenarios and modifiers.*

| Scenario | Deaths per 100,000 (± 1 standard deviation) | | | |
|---|---|---|---|---|
| | Lower-risk | Higher-risk (community) | Higher-risk (LTC residents) | Overall |
| **NS** (Fig. 2, col. 1) | 93.7 (90.3, 97.1) | 4702.1 (4615.9, 4788.3) | 4532.6 (4280.0, 4785.2) | 415.1 (408.5, 421.6) |
| **IS** (Fig. 2, col. 2) | 92.4 (89.4, 95.5) | 2090.9 (2038.0, 2143.8) | 613.6 (478.4, 748.8) | 221.7 (217.8, 225.5) |
| **PS** (Fig. 2, col. 3) | 92.5 (89.4, 95.7) | 18.6 (0.0, 38.3) | 54.3 (0.0, 123.3) | 87.6 (84.2, 91.1) |
| **NS + RC** (Fig. 3, col. 1) | 74.2 (71.4, 77.0) | 3712.8 (3633.0, 3792.6) | 3336.5 (3114.1, 3558.8) | 326.2 (319.8, 332.5) |
| **IS + RC** (Fig. 3, col. 2) | 80.3 (77.4, 83.3) | 3491.5 (3337.7, 3645.3) | 3783.7 (3516.2, 4051.2) | 321.2 (309.7, 332.7) |
| **PS + RC** (Fig. 3, col. 3) | 81.2 (78.2, 84.3) | 3114.4 (3010.8, 3218.0) | 3944.2 (3695.7, 4192.6) | 299.5 (292.0, 307.1) |

*Table S2: Deaths per 100,000 (normalised by group) for the main text scenarios. The darker the colour in a column, the higher the number of deaths.*

| Reference | 2011 Census | Report 9 (1) (* indicates interpolated data) | | |
|---|---|---|---|---|
| Age group | Population (%) | Symptomatic requiring hospital treatment (%) | Hospitalised requiring ICU support (%) | Symptomatic requiring ICU support (%) |
| 0-9 | 11.8 | 0.1 | 5.0 | 0.005 |
| 10-19 | 12.1 | 0.3 | 5.0 | 0.015 |
| 20-29 | 13.6 | 1.2 | 5.0 | 0.060 |
| 30-39 | 13.1 | 3.2 | 5.0 | 0.160 |
| 40-49 | 14.6 | 4.9 | 6.3 | 0.309 |
| 50-59 | 12.2 | 10.2 | 12.2 | 1.244 |
| 60-64 | 6.0 | 15.0* | 23.6* | 3.722* |
| 65-69 | 4.8 | 18.5* | 31.4* | 6.036* |
| 70-79 | 7.1 | 24.3 | 43.2 | 10.498 |
| 80+ | 4.7 | 27.4 | 70.9 | 19.356 |

*Table S3: The values used for the hospitalisation calculations. The population breakdown is from the 2011 census, while the data in the final three columns is taken from (1). An asterisk denotes data that has been interpolated.*